\documentclass[times,twocolumn,final]{elsarticle}

\usepackage{cnf}

\usepackage{framed,multirow}

\usepackage{amssymb}
\usepackage{latexsym}

\usepackage{url}
\usepackage{xcolor}
\definecolor{newcolor}{rgb}{.8,.349,.1}

\usepackage{hyperref}

\usepackage[switch,pagewise]{lineno} 

\usepackage[version=4]{mhchem}
\usepackage{chemformula}
\usepackage{mathrsfs}
\usepackage[per-mode=symbol]{siunitx}
\usepackage{upgreek}
\usepackage{verbatim}
\usepackage{fixmath}
\usepackage{lipsum}
\usepackage{tabularx}
\usepackage{amsmath,bm}
\usepackage{caption}
\usepackage{graphicx}
\usepackage{times}
\usepackage[pagewise]{lineno}
\usepackage{comment}
\usepackage{balance}
\usepackage{siunitx}
\journal{Preprint submitted to Combustion and Flame}

\begin{document}

\verso{Bao et al.}

\begin{frontmatter}

\title{Influence of an oxide cap on the combustion of aluminum particles using boundary-layer resolved simulations}%

\author[1]{Hesheng \snm{Bao}}

\author[1]{Hongchao \snm{Chu}\corref{cor2}}
\cortext[cor2]{Corresponding author: Hongchao Chu, Templergraben 64, Aachen, 52056, Germany.}
\emailauthor{h.chu@itv.rwth-aachen.de}{Hongchao Chu}  

\author[1]{Shibam \snm{Bose}}

\author[1]{Heinz \snm{Pitsch}}

\address[1]{Institute for Combustion Technology, RWTH Aachen University, Aachen, 52056, Germany}

\begin{abstract}
A distinctive feature of burning aluminum particles is the 
condensation of the primary combustion product, alumina, forming an oxide cap that covers part of the aluminum particle and alters the local combustion characteristics. This study systematically investigates the effects of an oxide cap on the combustion characteristics and interfacial processes of aluminum particles, using boundary-layer resolved simulations, considering variations in cap size, cap position relative to the flow direction, and the particle Reynolds number, $\mathrm{Re_p}$. 
The oxide cap modifies both the interfacial transfer and the surrounding combustion characteristics. Front caps
enhance local evaporation near the junction between the cap and the aluminum body, whereas heat absorption by rear caps contributes to lower reaction temperatures. Variations in integrated heat release generally
follow those in the evaporation rate. Based on the simulation results, the Lagrangian modeling
framework was first revised for spherical particles. The film state is approximated using a 0.55-rule applied to temperature, species mass fractions, and pressure, from which the remaining thermophysical properties are derived. A simplified composition scheme enables film-state viscosity and fuel diffusivity to be pre-calculated as functions of temperature. Optimized
Whitaker and Schiller-Naumann correlations were developed for the mass evaporation rate and drag force coefficient, respectively, showing good agreement with the simulation results. The framework was further extended to doublet particles with oxide caps. Accounting for the exposed aluminum area provides
good evaporation predictions for rear caps, while front caps require a correction to the surface-averaged evaporation mass flux, particularly at high $\mathrm{Re_p}$. A correction depending
on the film Reynolds number and the fraction of the aluminum surface covered by the oxide cap was proposed.
Using $A_{\rm c}=\pi r_1^2$ as the reference area,
where $r_1$ is the radius of the spherical aluminum body,
the optimized spherical drag correlation generally provides an approximate description of the capped-particle results.

\end{abstract}


\end{frontmatter}


\section*{Novelty and significance statement}
The key novelty of this study lies in the systematic analysis of how oxide caps, differing in size and position, affect the interfacial processes of burning aluminum particles over a range of particle Reynolds numbers. {These effects are {then included} in a revised Lagrangian modeling framework.} The method incorporates an appropriate approximation of the film state and simplified temperature-dependent correlations for the film-state viscosity and fuel diffusivity that can be pre-evaluated before Lagrangian simulations. {A correction accounting for the film Reynolds number and the fraction of the aluminum surface covered by the oxide cap improves evaporation predictions for front-cap configurations. The optimized spherical drag correlation approximately describes the capped-particle results using a reference area based on the aluminum-body radius.} These findings connect boundary-layer
resolved combustion characteristics with practical modeling of mass transfer and drag within the investigated parameter range. 

\section{Introduction}
\label{sec:introduction}
Metallic particles are promising energy {carriers} due to their carbon-free nature, high energy density, and ease of transport~\cite{bergthorson2015direct}. Aluminum (Al) is a promising inorganic fuel candidate, being the most abundant metallic element in the Earth's crust and exhibiting relatively high reactivity. Its primary combustion product, alumina (\ch{Al2O3}), acts as
a functional material in diverse industrial applications. {Moreover, a} reduction of alumina back into metal fuels is also possible using clean primary energy sources~\cite{bergthorson2018recyclable}, facilitating a sustainable closed-loop energy cycle. 

Experimental investigations on the combustion of a single micro-sized aluminum particle date back to the 1960s~\cite{friedman1962ignition}. It is observed that the aluminum particle in \ch{O2} dominated oxidizers initially burns in bright vapor-phase with
spherically symmetric envelopes, followed by an asymmetric combustion induced by the condensation of alumina on the particle surface, forming a doublet particle. The condensed alumina is referred to as the oxide cap, or lobe~\cite{braconnier2018detailed}. The cap grows as combustion proceeds, affecting the combustion behavior. Beckstead et al.{~\cite{beckstead2004summary}} summarized the experimental efforts on the combustion of single aluminum particles, and pointed out that the burning time is significantly affected by the oxide cap. {Various modeling approaches have been employed to describe the geometry of doublet particles~\cite{beckstead2004summary,gallier2021role,liu2021modeling}.}  Gallier et al.{~\cite{gallier2021role}} calculated the shape of the doublet particle using the force equilibrium at the triple {line} among the surrounding gas, aluminum, and alumina{, which allows detailed simulations of the boundary layer around a particle. It was found that the existence of the cap distorts the flame structure, bringing hot oxides {back to the cap. An intense thermophoretic }motion of nano-sized alumina droplets, referred to as the ``smoke'', towards the cap was observed.} More recently, several studies have included the cap structures in boundary-layer resolved simulations, focusing mainly on the formation of the nano-sized alumina smoke~\cite{finke2024unsteady,finke2024population}. {While the aforementioned studies employed nearly quiescent ambient flow, Gallier focused on the combustion behavior of aluminum particles in convective flows, and noted the potential effects of an oxide cap on the evaporation of an aluminum particle~\cite{gallier2023aluminum}. The detailed effects of such a cap in convective flows, however, have not been sufficiently investigated. The cap is potentially positioned both downstream and upstream of the aluminum particle. The consequence of the two configurations has not been adequately explored.}

{Although resolving the boundary layer provides sufficient details, many applications involve numerous aluminum particles, making such simulations computationally prohibitive. A more practical alternative is to treat the aluminum particles as Lagrangian particles, where, in general, interfacial heat, mass, and momentum transfer are described by {analytical} correlations formulated in terms of non-dimensional numbers that characterize the boundary layer~\cite{ranz1952evaporation,whitaker1972forced,schiller1933drag}. However, the interfacial characteristics are significantly influenced by chemical reactions, making the standard correlations not applicable. In the presence of large temperature gradients, different correlations have been proposed, depending on the fuel type and conditions~\cite{renksizbulut1983experimental,kurose2003effects,lee1981modeling}. This was attributed to variations in the Stefan flow~\cite{jayawickrama2019effect}, kinematic viscosity~\cite{nagata2018direct}, and reaction rates~\cite{zhang2020drag} in the vicinity of the particle. These raise the need for physical interpretation in the correlations. Nevertheless, a simple and practical solution is a case-dependent fitting of existing correlations using the overall evaporation rate or drag force~\cite{zhifu2013evaluation}. Recently,} Thijs et al.{~\cite{thijs2024improvement}} developed a readily applicable Lagrangian framework for modeling heat and mass transfer in iron particle combustion under weak convective flows. The applied correlation and strategy for characterizing the boundary layer differ from those traditionally adopted for liquid fuels~\cite{knudsen2015modeling}. {It is worth mentioning that the Lagrangian modeling of heat, mass, and momentum transfer has rarely been examined for non-spherical burning particles. Compared to iron particles, aluminum particle combustion involves non-spherical geometry due to the presence of an oxide cap, which alters the combustion dynamics and associated interfacial processes. }{Most research employing the Lagrangian framework for aluminum particle combustion applies standard Lagrangian methods derived for conventional spherical hydrocarbon fuel droplets, with mass transfer modified by the cap {by assuming that the exposed part of the aluminum particle is not affected}~\cite{liu2021modeling,zhang2022detailed,ding2025multistage,zhang5145395combustion}. Although Emelyanov et al.~\cite{emelyanov2020dynamics} have noticed that the existence of an oxide cap changes the drag force based on boundary-layer resolved simulations, this has not been accounted for in Lagrangian modeling.}

{Thus, the present study focuses on the detailed effects of an oxide cap for aluminum particles in convective flows and examines how the non-spherical cap geometry influences Lagrangian modeling}. Numerical simulations were performed by resolving the boundary layers around burning aluminum particles. {The effects of } oxide caps, positioned both upstream and downstream of the aluminum particle, with different sizes, were investigated over a range of Reynolds numbers. The resulting combustion characteristics were analyzed in detail. Based on the simulation results, the validity of the commonly used Lagrangian models and correlations was systematically assessed, {and a revised Lagrangian modeling method that incorporates an appropriate characterization of the boundary layer was proposed for aluminum particle combustion.} In particular, the correlations for mass transfer and drag force were revised for burning aluminum particles and extended to account for variations in cap size and position. The present work enhances the understanding regarding the effects of an oxide cap and provides a reliable foundation for Lagrangian modeling of burning aluminum particles.

\section{Numerical details and relevant theory}

\subsection{Particle and oxide cap shape}
\label{sec:geometry}
\begin{figure}[h!]
\centering
\includegraphics[width=192pt,trim=0pt 0pt 0pt 0pt, clip]{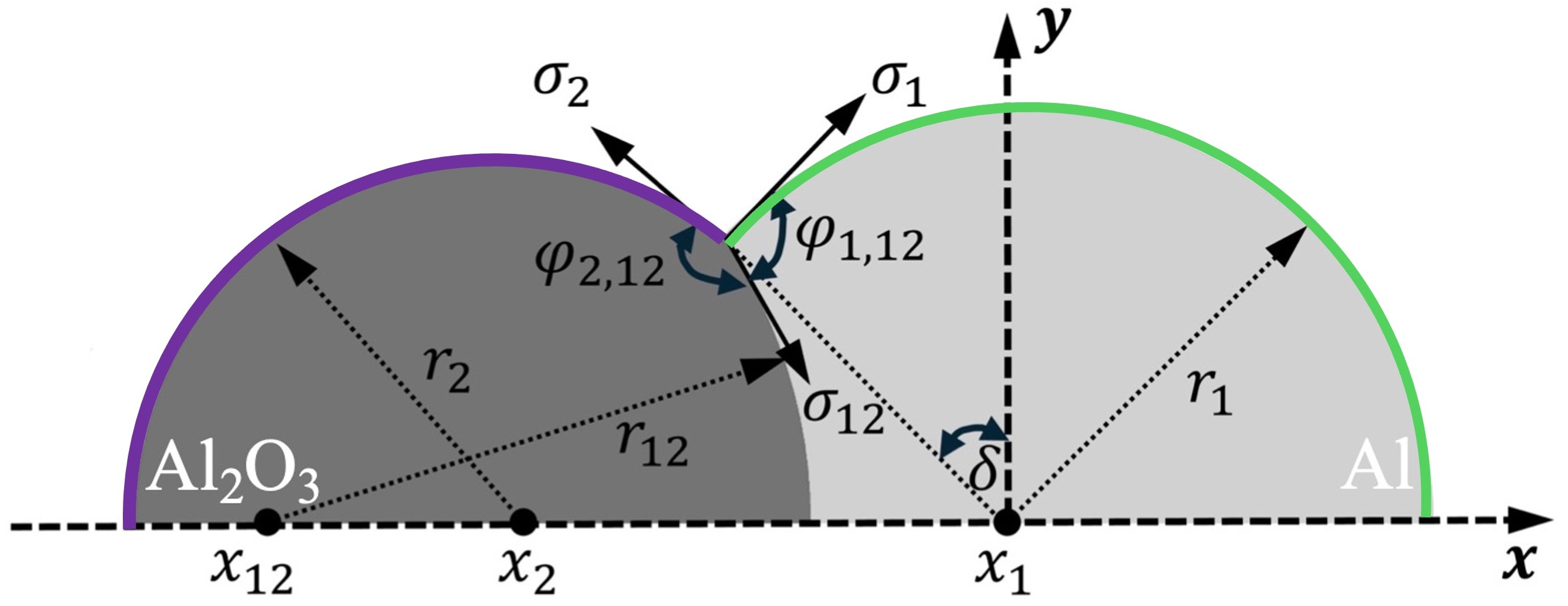}
\vspace{-10 pt}
\caption{\footnotesize Geometry of an aluminum particle with an oxide cap.}
\label{fig:lobe}
\vspace{-5 pt}
\end{figure}

In the combustion of aluminum particles, {the thin {passivating} alumina layer on the aluminum surface} is first heated above its melting temperature ($\thicksim$\SI{2330}{K}), and recedes into a protruding cap on the aluminum particle, forming a {non-spherical} doublet particle~\cite{braconnier2020experimental}. {As the combustion proceeds, the oxide cap grows, and the geometry of the doublet particle changes.} Figure~\ref{fig:lobe} schematically presents the geometry of the doublet particle by showing a slice that crosses its cylindrical symmetry axis. Detailed theory regarding the precise geometry, derived from the force balance at the triple {line} among aluminum, alumina, and the surrounding gas, can be found in Ref. \cite{gallier2021role}. Here, the subscript 1 refers to the aluminum and 2 refers to the alumina. The centers of the spheres $x_1$ and $x_2$, the radii $r_1$ and $r_2$, and the angle $\delta$ are case-dependent parameters used for determining the geometry in the simulations. The parameters are explicitly defined from the known mass of aluminum and oxide cap. The ratio of the aluminum particle surface covered by the oxide cap is $\beta_{\rm ox}=1-A_{\rm s, Al}/(4\pi r_1^2)=0.5(1-\sin\delta)$. Here, $A_{\rm s}$ is the surface area, and the subscript ``Al''  refers to the exposed aluminum body. Specifically, for a spherical particle without a cap, $\beta_{\rm ox}=0$. Note that, in this study, the particle diameter $d_{\rm p}$ in a doublet particle refers to the aluminum body, i.e., $d_{\rm p}=2r_1$, unless otherwise specified. Similarly, in the following, the aluminum particle refers to the non-spherical doublet particle system when an oxide cap is involved. The terms ``aluminum surface’’ and ``alumina surface’’ refer specifically to the {aluminum-gas interface and the alumina-gas interface} of the two respective bodies within the doublet particle, for brevity, and are indicated in the figure by {green and purple} colors, respectively.

\subsection{Boundary-layer resolved simulation}
This section presents the numerical details regarding the boundary-layer resolved simulations of a single burning aluminum particle, also referred to as a droplet. In this study, the simulations were performed by solving the governing equations in the gaseous phase around the aluminum particle, while the interfacial processes were represented through boundary conditions. Both spherical particles {without an oxide cap} and non-spherical particles with an oxide cap were considered. A two-dimensional axisymmetric computational domain was adopted. 

\begin{figure}[h!]
\centering
\includegraphics[width=192pt,trim=0pt 0pt 0pt 0pt, clip]{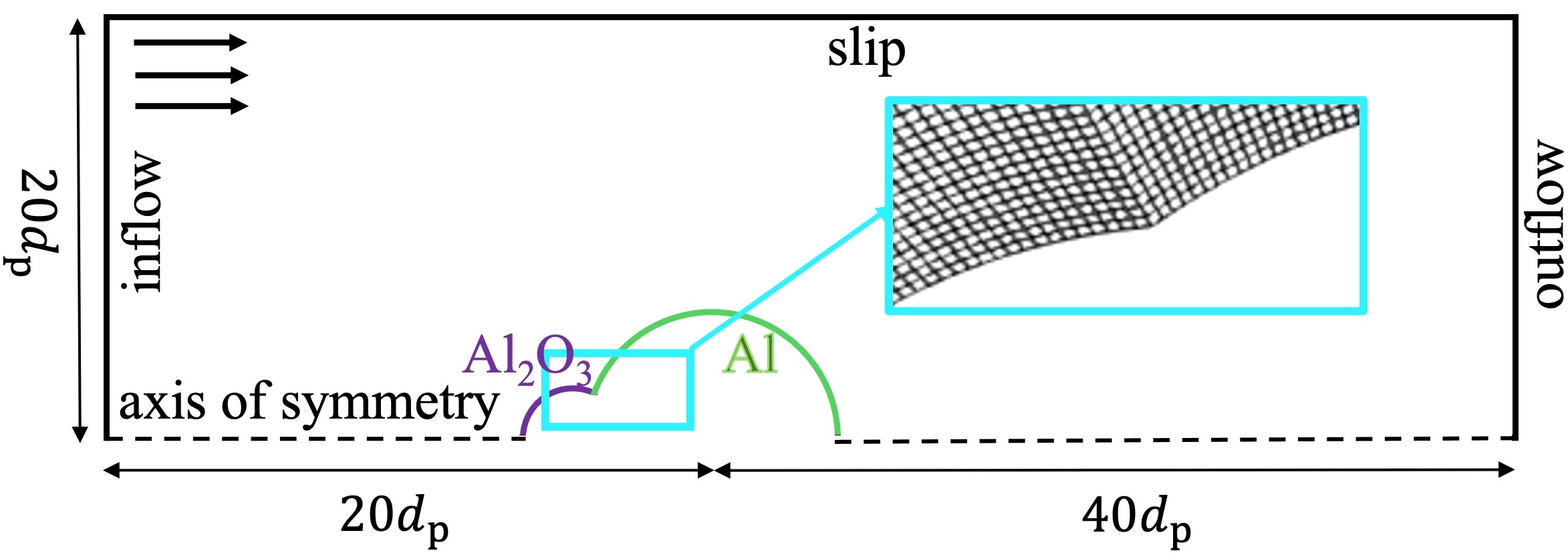}
\vspace{-10 pt}
\caption{\footnotesize Schematic illustration of the computational domain and applied boundary conditions. {Inset shows an} example of the zoomed-in mesh near the spherical boundaries.}
\label{fig:schematic}
\vspace{-5 pt}
\end{figure}
The simulations {were} performed using an extended version of the direct numerical simulation framework developed by Zhong et al.~\cite{zhong2020structure}, implemented in OpenFOAM using the finite-volume method~\cite{OF}. An implicit second-order backward differencing scheme {was} used for temporal integration. The timestep {was} dynamically adjusted during the simulation by fixing the Courant number at 0.35. Both the diffusion terms and the convection terms {were} discretized using a third-order cubic scheme. Given the emphasis on Lagrangian modeling of interfacial processes, the present study focuses exclusively on steady-state combustion. {The effects of geometry variations on interfacial characteristics were examined through simulations with different fixed geometries.} The simulations {were} terminated when the quasi-steady state {was} achieved. 

\subsubsection{Governing equations}
The governing equations for mass, momentum, species, and sensible enthalpy in  the gas phase read:
\begin{equation}
\label{eq:rho}
    \frac{\partial \rho}{\partial t} + \nabla\cdot(\rho \bm{u}) = 0,
\end{equation}
\begin{equation}
\label{eq:U}
    \frac{\partial {(\rho\bm{u})}}{\partial t} + {\nabla\cdot(\rho\bm{u}\bm{u})} = -\nabla p + \nabla\cdot\bm{\tau},
\end{equation}
\begin{equation}
\label{eq:Y}
     \frac{\partial {(\rho Y_i)}}{\partial t} + {\nabla\cdot(\rho\bm{u}Y_i)} = - \nabla\cdot \bm{J}_i +\dot{\omega}_i,
\end{equation}
\begin{equation}
\label{eq:h}
    \frac{\partial  {(\rho h_s)}}{\partial t} + {\nabla\cdot(\rho\bm{u}h_s)} = \frac{Dp}{Dt} -\nabla\cdot \bm{q} + \bm{\tau}:\nabla\bm{u}+\dot{\omega}_T,
\end{equation}
where {$\bm{\tau}$} is the viscous tensor. $\rho$, $p$, $Y$, $h_s$, $\bm{u}$ are the density, pressure, mass fraction, sensible enthalpy, and velocity vector, respectively. A subscript ``$i$" refers to species $i$.  $\bm{J}$ is a flux vector{, given by}\vspace{-4 pt}
\begin{equation}  
    \bm{J}_i=\rho(\bm{V}_{m,i} +\bm{V}_{T,i})Y_i. 
\end{equation}
$\bm{V}_{m,i}$ is the mass diffusion velocity described by the {Hirschfelder-Curtiss} approximation that is proportional to the mole fraction gradient, and includes a correction velocity $\bm{u}_c$ to ensure {local} mass conservation~\cite{poinsot2011theoretical}. $\bm{V}_{T,i}$ is the thermal diffusion velocity. 
The energy flux $\bm{q}$ involves the effect of heat conduction and diffusion of species:\vspace{-5 pt}
 \begin{equation}
    \bm{q}=-\lambda\nabla T + \rho\sum h_{s,i}Y_i(\bm{V}_{m,i}+\bm{V}_{T,i}),
\end{equation}
{$\lambda$ is the thermal conductivity and $T$ is the temperature}. $\dot{\omega}_i$ and $\dot{\omega}_T$ in Eqs.~(\ref{eq:Y}) and \ref{eq:h} are sources {due} to chemical reactions {and alumina condensation}. Radiation is neglected since its influence is minor~\cite{brooks1995dynamics}. This study focuses on the combustion of aluminum in the mixture of \ch{O2} and \ch{N2}. Thus, an adapted subset of 
the complete chemical mechanism from Glorian et al.~\cite{glorian2016role}, excluding species that involve \ch{Ar} and \ch{H}, is applied to calculate $\dot{\omega}_i$. The transport properties are based on standard kinetic gas theory, and the thermodynamic properties are described by the NASA polynomials. An ideal gas law is applied to calculate $\rho$ according to $p$, $Y_i${,} and $T$. Note that the size of liquid alumina smoke has been experimentally reported to be on the order of \SI{1}{\micro\meter} ~\cite{babuk2005burning,zarko2013formation}. In the currently applied mechanism, liquid alumina smoke is
described using gaseous transport equations. A ``gaseous'' \ch{Al2O3}$_{\rm ,L}$ that features the thermodynamic properties of the liquid \ch{Al2O3} is involved, by assuming an irreversible and fast condensation{, following ~\cite{gallier2021role,qiu2024detailed}}.

\subsubsection{Boundary conditions}
\label{subsubsec:bd}
{The exposed aluminum surface, indicated by Al in Fig.~\ref{fig:schematic}, separates the liquid aluminum body from the surrounding gas phase and is represented by a boundary.} Heat and mass transfer at the {surface {are}} described through a dynamically updated Dirichlet boundary condition for $Y_i$, $T$, and $\bm{u}$. The combination ensures interfacial mass and enthalpy balance. The species mass balance at the aluminum surface reads
\begin{equation}
\label{eq:Al_Y}
    (\rho \bm{u}Y_{i} + \rho( \bm{V}_{m,i}+\bm{V}_{T,i})Y_{i})
    \lvert_{\rm s}\cdot \bm{n}_{\rm s} = \dot{m}^{\prime\prime}_{i}.
\end{equation}
Here, the subscript ``s'' refers to the interface. It is noted that for an aluminum particle, the inclusion of differential diffusion in Eq.~(\ref{eq:Al_Y}), which is commonly neglected in existing research, is essential. The strong effects of differential diffusion are shown in Sec.~\ref{sec:results}. The Stefan flow velocity $\bm{u}_{\rm s}$ follows,
\begin{equation}\label{eq:Al_u}
    \rho \bm{u}_{\rm s}\cdot\bm{n}_{\rm s} = \sum\dot{m}^{\prime\prime}_{i} = \dot{m}^{\prime\prime},
\end{equation}
where $\bm{n}_{\rm s}$ is the surface unit normal vector that points outward from the computational domain. $\dot{m}^{\prime\prime}_{i}$ is {the mass flux due to evaporation or heterogeneous surface reactions}. In the current study, heterogeneous surface reactions are not considered. This simplification is acceptable since heterogeneous surface reactions are insignificant for the currently studied particle diameters~\cite{glorian2016role}. {The deposition of alumina is neglected here, following Gallier~\cite{gallier2023aluminum}. It is assumed that all other species are insoluble in liquid aluminum.} Evaporation is the only process that contributes to $\dot{m}^{\prime\prime}_{i}$.  Thus, $\dot{m}^{\prime\prime}_{i}$ is referred to as the surface evaporation rate. Since aluminum is the only species that evaporates, $\dot{m}^{\prime\prime}_{i}=\dot{m}^{\prime\prime}$ when $i=\ch{Al}$, and $\dot{m}^{\prime\prime}_{i}=0$ otherwise. Ignoring the radiative heat transfer, the balance of energy on the aluminum surface reads
\begin{equation}\label{eq:Al_h}
    -(\lambda \nabla T )\lvert_{\rm s}\cdot\bm{n}_{\rm s} = -\dot{m}^{\prime\prime}_{i}h_v,
\end{equation}
$h_v$ is the heat of vaporization, which is a function of $T$ and is taken from NIST Chemistry WebBook~\cite{nist}. Here, it is assumed that the interface is at local equilibrium. Accordingly, $Y_{\rm Al,s}$ is derived from Dalton's law of partial pressure:
\begin{equation}\label{eq:Al_p}
    Y_{\ch{Al}{\rm ,s}} = \frac{p_{\rm Al,sat}W_{\rm Al}}{(p_{\rm am}-p_{\rm Al,sat})W_{\rm am}+p_{{\rm Al},{\rm sat}}W_{{\rm Al}}},
\end{equation}
where the subscript ``sat'' refers to the saturation state, and the subscript ``am'' corresponds to the ambient gas. $W$ denotes the molar weight. The saturated vapor pressure of aluminum is a function of $T$, provided in Ref.~\cite{honig1962vapor}. At each timestep, Eqs.~(\ref{eq:Al_Y})-(\ref{eq:Al_p}) are iteratively solved for each computational boundary face discretizing the aluminum surface until a combination of $Y_i$, $T$, and $\dot{m}^{\prime\prime}$ is obtained that simultaneously satisfies both the mass and energy balances.  Using Eq.~(\ref{eq:Al_u}), the Stefan flow velocity is then imposed as the Dirichlet boundary condition for $\bm{u}$. 

Similar to the aluminum surface, the surface of the oxide cap is represented by a boundary {(\ch{Al2O3} in Fig.~\ref{fig:schematic})}. A no-slip boundary condition is applied. The temperature of the alumina surface is set to the temperature averaged over the exposed aluminum surface~\cite{gallier2023aluminum}. This assumption is considered valid since the Biot number for the currently studied doublet particle is very small. 

\subsubsection{Simulated cases and relevant details}
Experimental studies have shown that the formation of an oxide cap in diluted \ch{Ar}-dominated environments is suppressed~\cite{beckstead2004summary}. Thus, the ambient gas mixture is chosen as 21\% \ch{O2} and 79\% \ch{N2} in volume, with a temperature of \SI{300}{K} at a pressure of \SI{1}{atm}. {The main parametric study was conducted for a fixed aluminum mass corresponding to a spherical diameter of} $\SI{70}{\micro\meter}$, which is a typical size investigated in experiments~\cite{braconnier2020experimental,braconnier2021aluminum}. {An additional spherical particle with a diameter of \SI{210}{\micro\meter} was simulated for comparison with available measurements~\cite{bucher2000detailed}.} The effect of cap size was studied by varying the mass fractions of alumina in the non-spherical doublet particle, $f_{\rm ox}$, while conserving the mass of aluminum. Four {values of} $f_{\rm ox}$, 0 (no cap, spherical), 0.1, 0.3, and 0.7 were considered, corresponding to $\beta_{\rm ox}=0$, 0.09, 0.21 and 0.46, respectively{, following the idealized geometry proposed in Ref.~\cite{gallier2021role} and described in Sec.~\ref{sec:geometry}}. {Note that for $f_{\rm ox}=0.7$, an aluminum particle with a diameter of 70\,\si{\micro\meter} corresponds to $r_1=\SI{39.7}{\micro\meter}$.}

For each $f_{\rm ox}$, {simulations {were} conducted} for different particle Reynolds numbers ${\rm Re_p}=\rho_\infty u_\infty d_{\rm p}/\mu_\infty$. Note that $d_{\rm p}$ denotes the equivalent sphere diameter corresponding to the fixed aluminum mass, namely \SI{70}{\micro\meter}. $\mu$ is the viscosity. The subscript $\infty$ refers to the far-field. The Weber number, ${\rm We}=\rho_\infty u_\infty^2 d_{\rm p}/\sigma_{\rm Al}$, represents the ratio of inertial to surface tension force acting on a particle. When ${\rm We}<1$, the surface tension dominates, preventing particle deformation and breakup. For the currently studied conditions, the criterion ${\rm We}<0.5$ corresponds to a maximum ${\rm Re_p}\approx240$, assuming a particle temperature of \SI{3000}{K}. {The Mach number is about 0.15.} Therefore, ${\rm Re_p}$ values ranging from 0 (quiescent environment) to 200 were investigated. {It is worth noting that according to Emelyanov et al.~\cite{emelyanov2020dynamics}, the ${\rm Re_p}$ range considered in the present study lies within the regime where the flow around the doublet aluminum particle remains symmetric with respect to the meridional plane, avoiding the periodical oscillation of {the} wake for higher ${\rm Re_p}$. This justifies the application of a two-dimensional axisymmetric computational domain.} 

For each non-zero $f_{\rm ox}$, the oxide cap was positioned either upstream (front cap) or downstream (rear cap) relative to the flow. In the current study, a temperature of \SI{2497}{K} {was} adopted for all cases to compute {the surface tensions, following Ref.~\cite{gallier2021role}}. This value corresponds to the $T_{\rm s}$ of a perfectly spherical burning aluminum particle in a quiescent environment. This simplification is justified because a variation of \SI{200}{K} from the currently selected $T_{\rm s}$, which exceeds the maximum variation observed across the studied cases, results in less than a 2 \% change in $r_1$. With variations in $f_{\rm ox}$, $\rm Re_{p}$, and relative cap positions in the flow, {simulations of} more than 50 cases {were} conducted, each requiring 40 to 50 hours with 192 processors, depending on ${ \rm Re_p}$. {For the experimental comparison, the ambient conditions specified above were retained for the {\SI{210}{\micro\meter}} particle, with $\mathrm{Re_p}=6.38$. In addition, the front- and rear-cap cases at $f_{\rm ox}=0.3$ were considered. These {\SI{210}{\micro\meter}} cases were not used to calibrate the evaporation correlations.}

\subsection{Key properties derived from simulations}
\label{subsec:theory_resolved_particle}
As described in Sec.~\ref{subsubsec:bd}, the surface evaporation rate $\dot{m}^\prime\prime$ is evaluated locally at each aluminum boundary face of the computational cells. The average evaporation rate for the particle is then obtained by integrating over the {simulated} aluminum surface:
\begin{equation}\label{eq:Al_mprimeprime}
\langle \dot{m}^{\prime{\prime}}\rangle = \frac{1}{A_{\rm s,Al}}\int_{A_{\rm s,Al}}\dot{m}^{\prime{\prime}}, dA_{\rm s,Al},
\end{equation}
{where the integration is performed over the surface area of the aluminum surface, as described in Sec.~\ref{sec:geometry}}. In the following, $\langle\phi\rangle$ represents the surface-averaged value of a property $\phi$ over the aluminum surface. We emphasize that the overall evaporation rate, the mass evaporation rate $\dot{m}$, differs from the evaporation rate per unit area, $\langle \dot{m}^{\prime{\prime}} \rangle${: $\dot{m} = \langle \dot{m}^{\prime{\prime}} \rangle A_{\rm s,Al}$}.

The drag force $\bm{F}_{\rm D}$ contains two components: 
\begin{equation}\label{eq:drag}
    \bm{F}_{\rm D} = \bm{F}_{p}+\bm{F}_{\tau},
\end{equation} 
{with the pressure given by}
\begin{equation}\label{eq:drag_p}
    \bm{F}_{p} = \int_{A_{\rm s,tot}} (p_{\rm s}-p_{\rm \infty})\cdot \bm{n}_{\rm s}\ dA_{\rm s,tot},
\end{equation} 
{and the viscous force reads}
\begin{equation}\label{eq:drag_tau}
    \bm{F}_{\tau} = -\int_{A_{\rm s,tot}} \mu(\nabla\bm{u}+\nabla\bm{u}^{\rm{T}}-\frac{2}{3}(\nabla\cdot\bm{u}){\bm{I}})\cdot \bm{n}_{\rm s}\ dA_{\rm s,tot}.
\end{equation}
{Here, the integration is carried out over the aluminum surface and, if present, the alumina surface. $A_{\rm s,tot}$ refers to the area of both the aluminum and alumina surfaces.} Due to the axisymmetry, the force components normal to the symmetry axis are zero.


\subsection{Lagrangian approach}
\label{sec:lagrangian}
Practical applications involve multiple particles. To resolve the boundary layers of each particle is computationally unaffordable. {Therefore, the} concept of Lagrangian particles is preferred. In that case, models are used to describe the evaporation process, instead of resolving the boundary layer around a particle. {One of the main contributions of this study is the revision of the Lagrangian modeling method for aluminum particle combustion, using results from boundary-layer resolved simulations. The relevant Lagrangian theory is therefore summarized here, with a focus on modeling the mass transfer and the drag force.}

\subsubsection{Mass transfer and drag force for a spherical particle}
\label{subsec:equations_single}
{Within the Lagrangian framework, the mass evaporation rate of a perfectly spherical particle is evaluated using properties from a reference film state{, which are approximated using film rules and will be explained in detail later,} as follows~\cite{sirignano2010fluid}:}
\begin{equation}\label{eq:lagrangian_m}
    \dot{m} = \langle \dot{m}^{\prime{\prime}}\rangle A_{\rm s,Al} = \rho_{\rm f} D_{\rm f}  {\rm{{Sh}_{eff}}}/d_{\rm p}\cdot A_{\rm s,Al}.
\end{equation}
{Here $\rm Sh_{eff}$ is the effective Sherwood number, which is approximated using film properties that will be explained later.} $D$ represents the mass diffusivity of the fuel vapor. {For a spherical particle without a cap, $A_{\rm s,Al}=\pi d_{\rm p}^2$.} In Lagrangian simulations, the mass of a particle $m_{\rm p}$ is tracked, and $d_{\rm p} = (6m_{\rm p}/(\rho_{\rm p}\pi))^{1/3}$ {for a spherical particle}. Note that the subscript ``f'' refers to a reference film state, corresponding to the characteristics of the shear layer around the particle.  The film state must be delicately chosen to correctly approximate the characteristics of the boundary layer around a particle. For a selected film quantity $\phi$, it can be determined using a film rule ($\alpha$-rule) that is a weighted average of the far-field ``$\infty$'' and  the surface: 
\begin{equation}\label{eq:lagrangian_rule}
    \phi_{\rm f} = \alpha\phi_\infty + (1-\alpha)\phi_{\rm{s}}.
\end{equation}
$\alpha$ is the rule factor to be varied. Different values of $\alpha$ {have} been used in the literature. Whitaker has applied the ambient temperature ($\alpha =1$) to characterize the film~\cite{whitaker1972forced}, while Miller et al.{~\cite{miller1998evaluation}} adopted the boiling state ($\alpha =0$) to model the evaporation of various liquids.  Intermediate factors, 1/2~\cite{xu2024phase,thijs2024improvement} and 1/3~\cite{hubbard1975droplet,knudsen2015modeling} have been applied. Knudsen et al.{~\cite{knudsen2015modeling}} noticed that the preferred rule factor depends on the type of fuel and the ambient temperature, and suggested 1/3 for liquid fuels. On the contra{ry}, Thijs et al.{~\cite{thijs2024improvement}} recommended 1/2 for the modeling of iron particles. In practice, the film rule can be applied either directly to the desired quantity~\cite{gallier2023aluminum,miller1998evaluation} or to related quantities from which the desired quantity is derived~\cite{law1972kinetics}. The Sherwood number, $\rm Sh$, quantifies the ratio of convective to diffusive mass transfer. The effective Sherwood number $\rm Sh_{eff}$, which accounts for corrections due to slip and Stefan flow effects for burning particles~\cite{thijs2024improvement,sirignano2010fluid}, is given by
\begin{equation}\label{eq:lagrangian_sh}
     {\rm{{Sh}_{eff}}} = {\rm Sh}\cdot{\rm{ln}}(1+{{B_{\rm{M}}}})/{{B_{\rm{M}}}}.
\end{equation}
${B_{\rm{M}}}$ is the Spalding mass transfer number:
\begin{equation}\label{eq:lagrangian_B}
{B_{\rm{M}}} = (Y_{{\rm{Al}},\rm{s}}-Y_{{\rm{Al}},\infty})/(1-Y_{{\rm{Al}},\rm{s}}).
\end{equation}
It is noted that ${\rm Sh}$ is obtained from empirically calibrated correlations, which are typically algebraic functions of characteristic film properties, the Reynolds number $\mathrm {Re_{f}}=\rho_{\rm f} u_\infty d_{\rm p}/\mu_{\rm f}$, Schmidt number $\mathrm{Sc_{f}}=\mu_{\rm f}/(\rho_{\rm f}D_{\rm f})$, and Prandtl number $\mathrm{Pr_{f}} = \mu_{\rm f}c_{p,\rm f}/\lambda_{\rm f}$. Here, $c_p$ and $\lambda$ refer to the specific heat and thermal conductivity, respectively. Therefore, the accuracy of modeling $\dot{m}$ depends on the choice of both the $\rm Sh$ correlation and the application of film rules to individual properties. {Note that ${\rm Re_f}$ refers to the film state, characterizing the shear layer around the particle, and should not be confused with ${\rm Re_p}${, which is defined using far-field quantities.}

In the current study, two commonly used {analytical} correlations for $\rm Sh$ were evaluated, namely the Ranz-Marshall correlation~\cite{ranz1952evaporation} and the Whitaker correlation~\cite{whitaker1972forced}, 
\begin{equation}
\label{eq:ranz}
    \mathrm{Sh} = 2+a\,\rm{Re_f}^{1/2}\rm{Sc_f}^{1/3},
\end{equation}
and 
\begin{equation}
\label{eq:whitaker}
    \mathrm{Sh} = 2+\mathrm{{Sc_f}^{0.4}}(b_1\,\mathrm{Re_f}^{1/2}+b_2\,\mathrm{Re_f}^{2/3}),
\end{equation}
respectively. Here, $a=0.6$, $b_1=0.4$, and $b_2=0.06$ are calibrated model constants. When $a=0.552$, Eq.~(\ref{eq:ranz}) becomes the Fr{\"o}sling correlation~\cite{frossling1958theoretical}. An appropriate Sh correlation for the combustion of aluminum particles, together with a proper film rule specific to individual properties, is essential. Therefore, several strategies for applying the film rule in Eq.~(\ref{eq:lagrangian_rule}) to identify $\rm Re_f$ and $\rm Sc_f$, and the associated film properties, were evaluated:
\begin{enumerate}
    \item applying the film rule to $T_{\rm f}$, $Y_{i,\rm f}$, and $p_{\rm f}$, and calculating $D_{{\rm f}}$, $\mu_{\rm f}$, and $\rho_{\rm f}$ from these properties, 
    
    \item directly applying the film rule to $D_{{\rm f}}$, $\mu_{\rm f}$, and $\rho_{\rm f}$,
    
    \item a combination of 1 and 2 for individual properties.
\end{enumerate}

The drag force is modeled using a coefficient $\rm{C_D}$ and the frontal area of the particle that is exposed to the flow, $A_{\rm c}$:
\begin{equation}\label{eq:lagrangian_drag}
\bm{F}_{\rm{D}} = \frac{1}{2} \rho_\infty{\rm{C_D}}  \bm{u}_\infty\lvert \bm{u}_\infty\rvert A_{\rm c}.
\end{equation}
$A_{\rm c}=\pi d_{\rm p}^2/4$. $\rm{C_D}$ is computed using calibrated correlations. The classical drag coefficient model, proposed by Schiller and Naumann~\cite{schiller1933drag}, is considered in the present study:
\begin{equation}\label{eq:Cd}
\mathrm{C_D} = \frac{24}{\mathrm{Re_f}}(c_1+c_2\,\mathrm{Re_f}^{c_3}),
\end{equation}
where $c_1 = 1$, $c_2 = 0.15$, and $c_3 = 0.687$ are calibrated constants, and need to be verified in the context of aluminum particles. 

\subsubsection{Mass transfer and drag force for a doublet particle with a cap}

When an oxide cap is present, a correction factor $\zeta$ is {employed in} Eq.~\ref{eq:lagrangian_m} to represent the variation in $\dot{m}$: 
\begin{equation}\label{eq:lagrangian_m_modified}
    \dot{m} =  \zeta\langle \dot{m}^{\prime\prime}\rangle A_{\rm s,Al} = \zeta \rho_{\rm f} D_{\rm f}  {\rm{{Sh}_{eff}}}/d_{\rm p}\cdot A_{\rm s,Al}.
\end{equation}
It is commonly assumed that the mass flux $\langle \dot{m}^{\prime\prime}\rangle$ remains the same as that without an oxide cap~\cite{zhang2022detailed,ding2025multistage}{. $A_{\rm s,Al} = (1-\beta_{\rm ox})\pi d_{\rm p}^2$ when an oxide cap is involved.} {Zhang et al.~\cite{zhang2022detailed} adopted the doublet particle geometry proposed by King~\cite{king2009aluminum}, in which the aluminum body remains spherical and is covered by an oxide cap. The particle diameter was defined as $d_{\rm p} = (6m_{\rm p,Al}/(\rho_{\rm p,Al}\pi))^{1/3}$, where $m_{\rm p,Al}$ is the aluminum mass in the doublet particle, and this diameter was used in the calculation of $A_{\rm s,Al}$ and $A_{\mathrm{c}}$.} {It is noted that, for consistency, the current paper adopts the diameter of the deformed aluminum particle shown in Fig.~\ref{fig:lobe}, when encountered, as $d_{\rm p}=2r_1$.}  The validity of assuming an $\langle \dot{m}^{\prime\prime}\rangle$ that is not affected by the presence of an oxide cap has not been justified in the literature. In addition, to the best of the authors’ knowledge, the drag force{, calculated from the $\rm C_D$ correlation following Eq.~\ref{eq:lagrangian_drag},} has not been revised for doublet aluminum particles with an oxide cap. {These are addressed in Sec.~\ref{sec:results}.}

\section{Results}
\label{sec:results}
{This section first {evaluates the boundary-layer resolved simulations against experimental measurements, followed by a detailed analysis of the effects of an oxide cap on the reacting boundary layer.} A revised Lagrangian method for predicting the mass transfer and drag force of burning aluminum particles, with and without an oxide cap, is established through comparisons with the simulation results.} 

\subsection{Global combustion characteristics}
{
\subsubsection{Validation of computational framework}
\label{sec:}
\begin{figure}[htp!]
\centering
\includegraphics[width=\columnwidth]{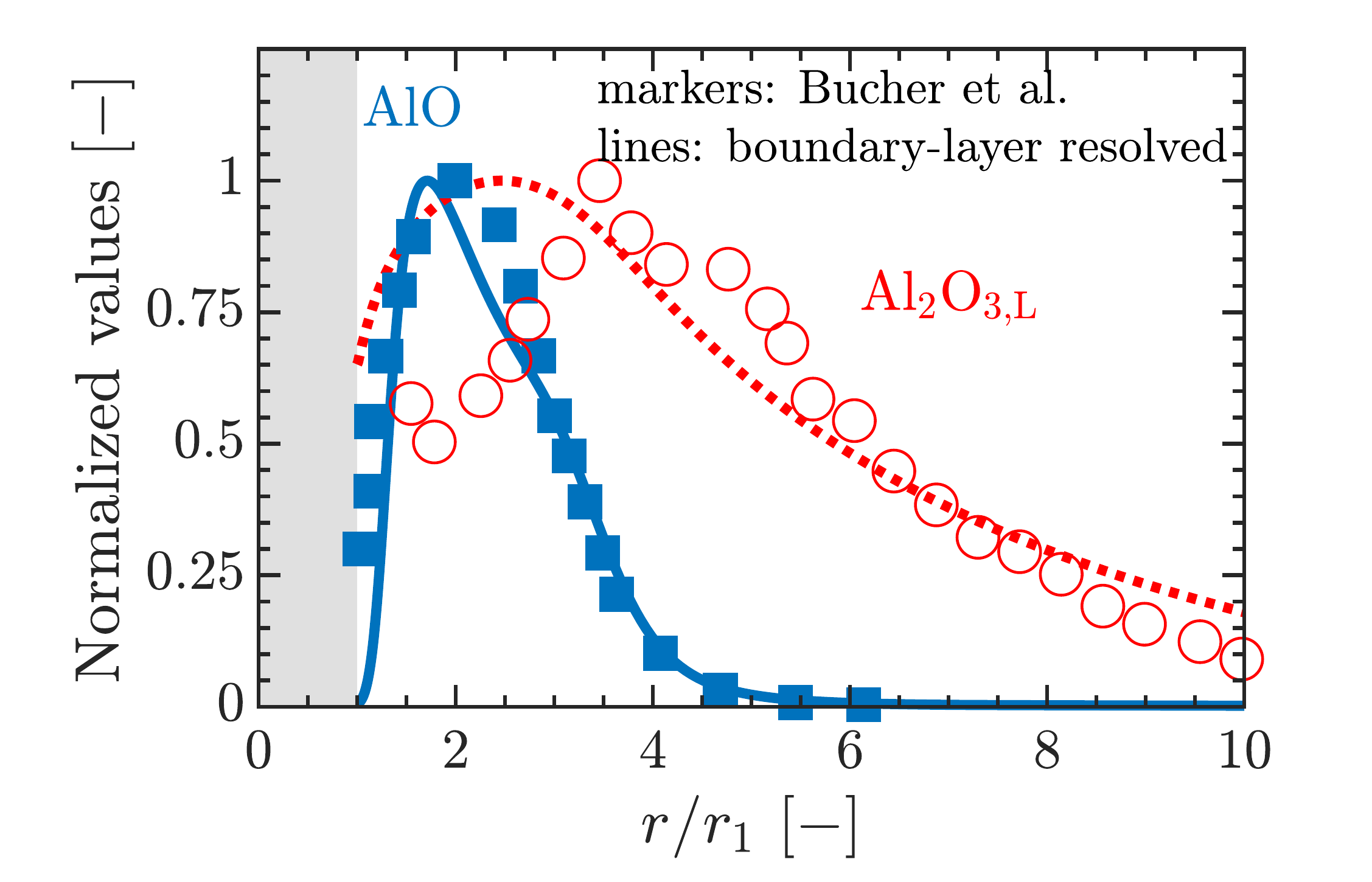}
\vspace{-10 pt}
\caption{\footnotesize Comparison with the measurements of \ce{AlO} (blue) and condensed \ce{Al2O3} (red) by Bucher et al.~\cite{bucher2000detailed}. Each experimental signal and time-averaged simulated mole-fraction profile is normalized by its own maximum. The gray region denotes the particle.}
\label{fig:bucher_validation}
\vspace{-10 pt}
\end{figure}

Figure~\ref{fig:bucher_validation} compares the normalized radial distributions of \ce{AlO} and condensed \ce{Al2O3} concentration with the measurements of Bucher et al.~\cite{bucher2000detailed}, plotted against $r/r_1$, where $r$ is the distance from the particle center and $r_1=\SI{105}{\micro\meter}$ is the particle radius. The simulations capture the \ce{AlO} peak near the particle and the broader distribution of condensed alumina farther from the surface. 
}

\subsubsection{Temperature distribution}
\label{sebsec:results:noCap}
\begin{figure*}[h!]
\centering
\includegraphics[width=395pt,trim=0pt 0pt 0pt 0pt, clip]{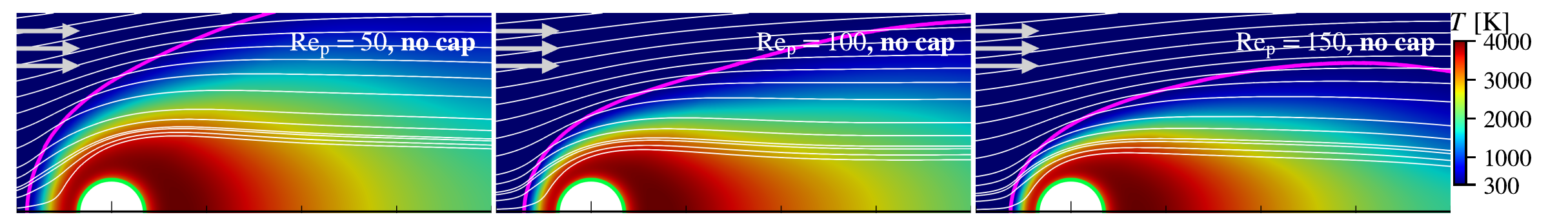}
\vspace{-5 pt}
\caption{\footnotesize Spatial distribution of temperature for perfectly spherical aluminum particles. White solid lines are streamlines. Magenta lines correspond to $Z_{\rm Al}=0.001$.}
\vspace{-5 pt}
\label{fig:contour_sphere}
\end{figure*}

\begin{figure*}[ht!]
\centering
\includegraphics[width=395pt,trim=0pt 0pt 0pt 0pt, clip]{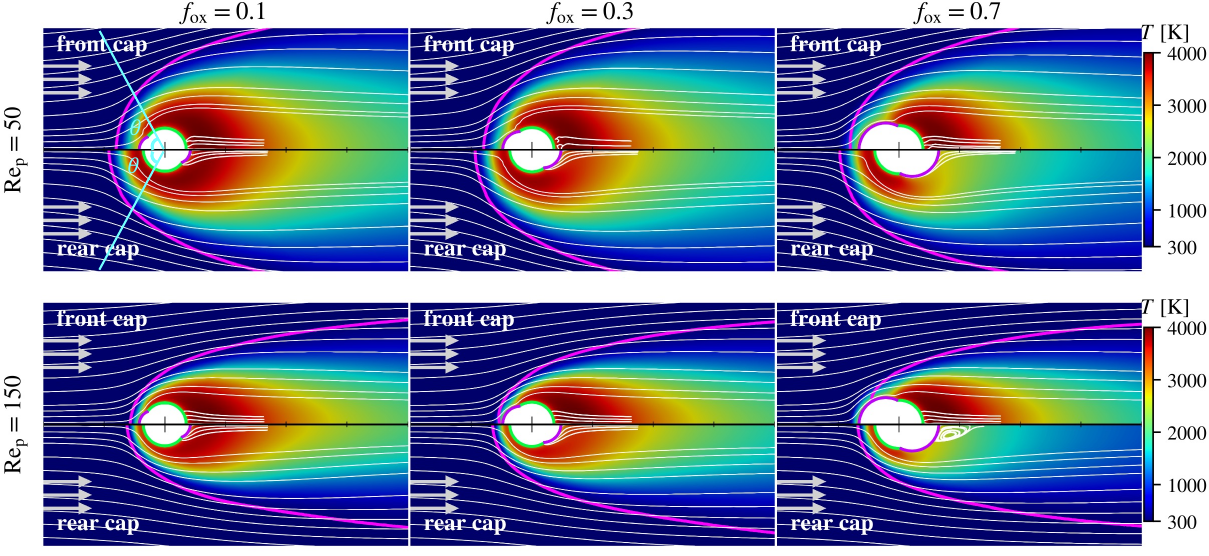}
\vspace{-5 pt}
\caption{\footnotesize Spatial distribution of temperature for aluminum particles with an oxide cap, in terms of different $\rm Re_p$, $f_{\rm ox}$, and relative positions of aluminum and alumina particles in the flow. White solid lines are streamlines. The aluminum surface and alumina surface are marked by green and purple colors. The definition of ``$\theta$'' is indicated for the front- and rear-cap configurations, respectively, in the plots of $f_{\rm ox}=0.1$ and $\rm Re_p = 50$.}
\vspace{-10 pt}
\label{fig:contour_cap}
\end{figure*}

Figure~\ref{fig:contour_sphere} presents the boundary-layer flow around perfectly spherical aluminum particles for selected $\rm Re_p$. The peak temperature appears at a distance from the aluminum surface, indicating a detached diffusion flame. This phenomenon is referred to as the standoff characteristic in the combustion of micro-sized aluminum particles~\cite{sundaram2016general}. As ${\rm Re_p}$ increases, the thickness of the boundary layer, indicated by the $Z_{\rm Al} = 0.001$ contour shown in magenta color, decreases. Here, $Z_{\rm Al}$ is the element mass fraction of Al, following $Z_{\rm Al}\coloneqq\,\sum {\epsilon}_i Y_i$, where ${\epsilon}_i$ represents the mass fraction of \ch{Al} in the $i$-th species. $Z_{\rm Al}$ serves as an indicator of the mixing state, with $Z_{\rm Al}=1$ referring to the fuel and $Z_{\rm Al}=0$ corresponding to the oxidizer. 

The effects of an oxide cap are analyzed by varying its size, its relative position with respect to the flow (upstream or downstream), and different $\rm Re_p$. The effects of these variations are illustrated in Fig.~\ref{fig:contour_cap} using representative cases. The boundary layer contracts toward the axis of symmetry with increasing $\rm Re_p$, which is consistent with the observations from the simulations of spherical aluminum particles. {The oxide cap substantially modifies the temperature field, particularly for the larger caps at higher $\mathrm{Re_p}$. For $f_{\rm ox}=0.3$ and $0.7$, the rear-cap configurations exhibit markedly lower peak temperatures than the corresponding front-cap configurations. The rear cap absorbs heat from the surrounding gas, while the front cap can supply heat to the gas at high $\mathrm{Re_p}$. This cap-position-dependent heat exchange contributes to the temperature differences observed in Fig.~\ref{fig:contour_cap}, and leads to the increased temperature observed for $Z_{\mathrm{Al}}<0.001$ in the upstream region of the front-positioned oxide cap when $f_{\rm ox}=0.7$ and $\mathrm{Re_p}=150$. Note that at high $\mathrm{Re_p}$, a local recirculation region appears downstream of the rear-positioned cap. This orientation dependence is consistent with the high-$\mathrm{Re_f}$ case studied by Gallier~\cite{gallier2023aluminum}. 
In the context of aluminum particles subjected to shocks,
Das and Udaykumar~\cite{das2021sharp} reported that the low-velocity recirculation region provides sufficient residence time for aluminum vapor to mix and react with oxygen. In the present study, however, the region of reversed flow exhibits weak heat release and does not coincide with the main reaction zone.}

{
\subsubsection{Mixing and heat release characteristics}
\label{subsubsec:mixing} 
In this section, the flame characteristics are evaluated within the region defined by the $Z_{\rm Al}=0.001$ contour, with an additional restriction of $r/d_{\rm p}\le10$ to exclude the far-downstream mixture from the statistics. Here, $r$ denotes the distance from the center of the aluminum body. For all cases, the currently defined flame zone captures all of the heat release. {The heat source} $\dot{\omega}_{T}$ considered here includes the contribution from alumina condensation and gas-phase chemical reactions. The former accounts for 46 to 50 \% of the volume-integrated $\dot{\omega}_{T}$ over the flame zone, referred to as $\dot{Q}$, for most of the simulated cases, consistent with the statement by Storozhev and Yermakov~\cite{storozhev2014interplay} that the latent heat released by \ce{Al2O3} condensation is comparable to the chemical heat release. Overall, $\dot{Q}$ increases with $\mathrm{Re_p}$, and decreases with $f_\mathrm{ox}$. At fixed $\mathrm{Re_p}$ and $f_\mathrm{ox}$, the rear cap generally yields a lower $\dot{Q}$ than the front cap. Note that variations in $\dot{Q}$ closely follow those in $\dot{m}$. For a spherical particle, $\dot{Q}$ increases from \SI{1.185}{\watt} at $\mathrm{Re_p}=0$ to \SI{2.249}{\watt} at $\mathrm{Re_p}=150$, corresponding to $\dot{m}$ of $2.800\times10^{-8}$\,\si{\kilogram\per\second} and $5.451\times10^{-8}$\,\si{\kilogram\per\second}, respectively. 
{Both the gas-phase reaction and alumina condensation contributions to $\dot{Q}$ generally follow $\dot{m}$. The differences in $\dot{Q}$ between the front- and rear-cap configurations are generally reduced when normalized by $\dot{m}$.} This highlights the importance of the evaporation modeling discussed in Sec.~\ref{subsec:mdot_sphere}. Only the rear-cap configuration at $f_{\rm ox}=0.7$ and $\mathrm{Re_p}=150$ exhibits a pronounced reduction in $\dot{Q}$ per unit $\dot{m}$. Compared to the front-cap configuration, reduced \ce{Al2O3} condensation accounts for approximately 74 \% of the difference in the reduced $\dot{Q}$. This reduction is associated primarily with 
a suppressed formation of \ce{Al2O3} via \ce{Al2O2} and \ce{O}. 

Figure~\ref{fig:angular_statistics} evaluates the statistics of selected resolved boundary layers within the defined region as functions of $\theta$. Here, $\theta=0^\circ$ corresponds to the upstream flow and $\theta=180^\circ$ to the downstream flow for all configurations. The solid lines show the density-weighted mean $Z_{\rm Al}$, are colored by the corresponding mean $T$ as indicated by the color bar, and refer to the black left-hand-side axis. The dotted lines show the density-weighted mean scalar dissipation rate $\chi$, referring to the yellow axis on the left-most side. {The outlined bars show the mass ratio density with respect to $\theta$, with the contribution from the heat-release region shown by the blue bars. Both refer to the right-hand-side axis. The area of each bar represents the mixture mass in that $\theta$-range relative to the total mixture mass of the selected region of the $f_{\rm ox}=0$, $\mathrm{Re_p}=0$ reference case. The integral of each distribution therefore corresponds to the mixture mass ratio. For the reference case, the mass ratio density reduces to the mass-weighted probability density function (PDF) of $\theta$ and integrates to unity, as shown by the dashed curves.}
The mixture mass retained in the selected region decreases with $\mathrm{Re_p}$, with a more concentrated distribution in the downstream area, accompanied by higher $T$ for the spherical particle. Compared to the spherical particle cases, the existence of a front cap significantly restricts the mixture formation in the upstream region, while the rear-cap configuration shows a reduced influence on {the shapes of the mass ratio density} and $Z_{\mathrm{Al}}$. However, relative to the front-cap case, the fraction of mixture mass within the heat-release region decreases markedly for the rear-cap cases. This is partially attributed to the absorption of heat by the rear cap, which leads to weakened reactions and reduced $T$. $\chi$ generally increases with $\mathrm{Re_p}$ for all configurations. For the spherical particle and the rear-cap case, $\chi$ peaks in the upstream region, while the front-cap case exhibits a peak near the junction between the cap and the aluminum body, indicating varied mixing characteristics. In contrast, the rear cap largely preserves the upstream mixing characteristics of the spherical particle, but reduces $\chi$ in the downstream region. The latter may also contribute to the weakened chemical reactivity due to the insufficient mixing between reactants and oxidizer.

\begin{figure*}[t]
    \centering
    \includegraphics[width=\textwidth]
    {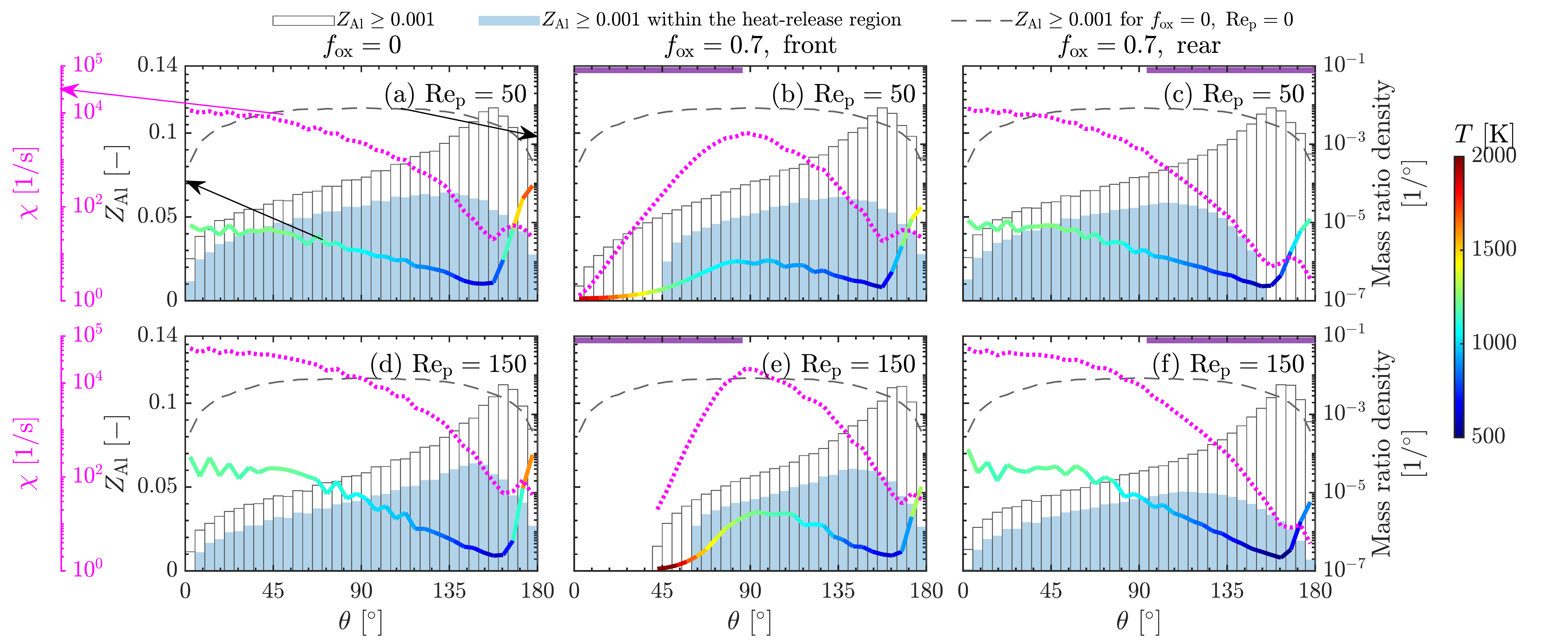}
    \vspace{-10 pt}
    \caption{
    Flame statistics within the region defined by  $Z_{\rm Al}\ge0.001$ and $r/d_{\rm p}\le10$ as functions of $\theta$, for selected cases. 
    {Outlined bars show the mass ratio density with respect to $\theta$, and blue bars show the contribution from the heat-release region. Their integrals refer to the corresponding mixture masses relative to the total mixture mass within the selected region of the $f_{\rm ox}=0$, $\mathrm{Re_p}=0$ reference case. The gray dashed curve shows the mass-weighted PDF of $\theta$ for the reference case and integrates to unity.}
    Colored curves show the density-weighted mean $Z_{\rm Al}$, with color indicating the corresponding density-weighted mean $T$. {Magenta} dotted curves show the density-weighted mean $\chi$. Purple bars indicate the $\theta$-range of the oxide cap on the particle surface.
}
\vspace{-5 pt}
    \label{fig:angular_statistics}
\end{figure*}
}

\subsection{Lagrangian modeling of mass transfer}
This section develops mass transfer models for aluminum particles. Different strategies for applying the film rule were employed, described in Section~\ref{subsec:equations_single}, with several $\rm Sh$ correlations. The predictions are evaluated by comparing them with the simulations of spherical particles. The {resulting} Lagrangian modeling method is further applied to non-spherical doublet particles with an oxide cap. A revised mass transfer model, suitable for different cap configurations, is proposed.

\subsubsection{Mass transfer modeling without oxide cap}
\label{subsec:mdot_sphere}
\begin{figure}[h!]
\centering
\includegraphics[width=185pt,trim=0pt 0pt 0pt 0pt, clip]{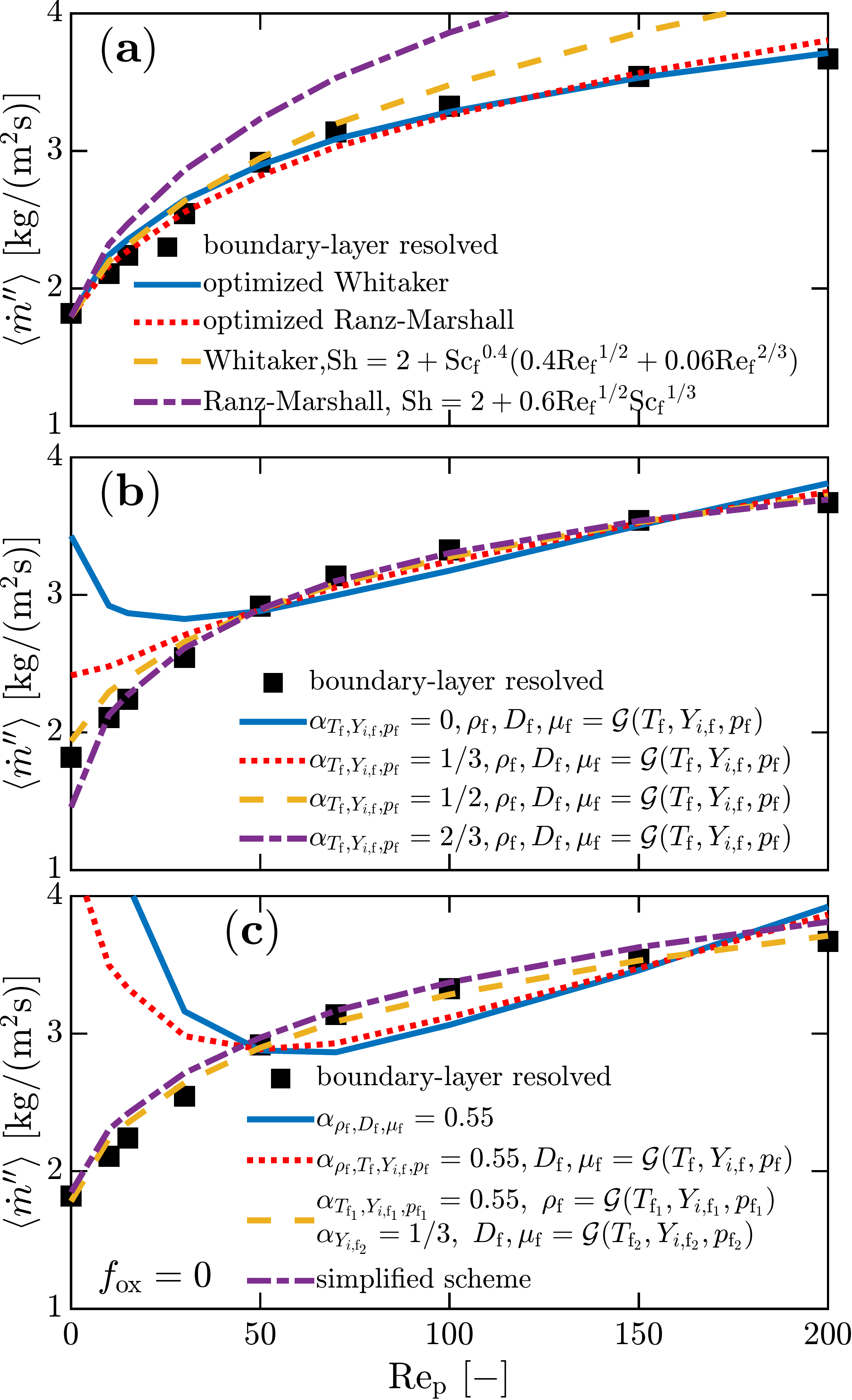}
\vspace{-5 pt}
\caption{\footnotesize Comparison of {boundary-layer resolved} surface-averaged evaporation mass flux and predictions using Lagrangian correlations: effects of (a) $\rm Sh$ correlations, (b) rule factors, and (c) strategies for applying rule factors to approximate the film state.}
\vspace{-10 pt}
\label{fig:calibration}
\end{figure}

This section evaluates the averaged surface evaporation rate $\langle \dot{m}^{\prime\prime}\rangle$ for spherical aluminum particles, corresponding to different $\rm Re_p$. Figure~\ref{fig:calibration}(a) presents the performance of the Ranz-Marshall and the Whitaker correlations, as well as their optimized versions. Here, strategy 1{, as described in Sec.~\ref{subsec:equations_single},} was adopted in combination with a rule factor of 0.55. The combined modeling method was selected by evaluating the prediction of $\langle \dot{m}^{\prime\prime}\rangle$ at $\rm Re_p=0$, a condition that yields ${\rm Sh}=2$ based on theoretical considerations. Using the chosen modeling method, the predicted $\langle \dot{m}^{\prime\prime} \rangle$ agrees well with the results of the boundary-layer resolved simulations, which are indicated by markers. Note that the surface-averaged properties, as defined in Eq.~\ref{eq:Al_mprimeprime}, are adopted. An increase in ${\rm Re_p}$ leads to a higher $\langle \dot{m}^{\prime\prime} \rangle$. It is seen that both the original Ranz-Marshall {correlation}, calibrated for water droplets in air, and the Whitaker {correlation}, which included oil droplets, overpredict $\langle \dot{m}^{\prime\prime}\rangle$. This is attributed to the combustion involved in the boundary layer. Optimized correlations are {hence required} for aluminum particles. The revised constant for the Ranz–Marshall correlation is $a=0.42$, while the optimized constants for the Whitaker correlation are $b_1=0.67$, $b_2=-0.15$. The former slightly underestimates $\langle \dot{m}^{\prime\prime}\rangle$. More complex correlations, such as {those} proposed in \cite{richter2012drag,clift2005bubbles}, are not suitable for the currently studied burning aluminum particles, and are not shown here for brevity.

The effects of the selected rule factor and the application of representative modeling strategies in calculating the film state are shown in Fig.~\ref{fig:calibration}(b) and (c), respectively. {Subscripts on $\alpha$ indicate the properties to which the film rule is applied, and $\mathcal{G}$ denotes the calculation of a property from the specified properties.} The results in Fig.~\ref{fig:calibration}(b) were obtained using the optimized Whitaker correlation using strategy 1. $\alpha=0$ (the surface state) and $\alpha=1/3$ were examined. However, predictions for both cases deviate substantially from the boundary-layer resolved simulations for $\rm Re_p<50$, which are common conditions for the combustion of aluminum particles. The suggestion of $\alpha=1/2$ by Thijs et al.~\cite{thijs2024improvement} in the simulation of iron particle combustion reasonably captures the values predicted by boundary-layer resolved simulations. A detailed analysis of different modeling strategies for individual film properties was conducted, with representative examples shown in Fig.~\ref{fig:calibration}(c). 
{It is concluded that directly applying a film rule to $\rho_{\rm f}$ results in substantial deviations from the boundary-layer resolved simulations. $D_{\rm f}$ and $\mu_{\rm f}$, on the other hand, can be evaluated with a different $\alpha$-rule applied to composition, giving similar evaporation predictions.}
Further analysis has indicated, but is not shown here for brevity, that the determination of $\mu_{\rm f}$ using strategies 1 and 2 shows negligible influence on the calculation of $\langle \dot{m}^{\prime\prime}\rangle$. This enables a simplified scheme for the Lagrangian modeling of aluminum particle combustion. The surface species composition is assumed to be a mixture of aluminum vapor and the ambient gas, where $Y_{\rm Al,s}$ is a temperature-dependent function following Eq.~\ref{eq:Al_p}. Based on this assumed surface composition, $D_{\rm f}$ and $\mu_{\rm f}$ can be calculated from the film state. In this way, $D_{\rm f}$ and $\mu_{\rm f}$ become functions of temperature only and can be pre-calculated and fitted before performing the Lagrangian simulation, thereby significantly reducing the computational cost in evaluating film transport properties when many particles are involved. {The performance of this simplified scheme is shown by the dashed-dotted line in Fig.~8(c).}

The present method retains the Lagrangian framework commonly used for hydrocarbon fuels~\cite{lee1981modeling,knudsen2015modeling}, {but uses film-state approximations and correlation
coefficients optimized for aluminum combustion}. Gallier et al.~{\cite{gallier2011combustion,gallier2023aluminum}} have demonstrated that $\dot{m}^{\prime\prime}$ for burning spherical aluminum particles can be accurately predicted by using properties from ``flame conditions", corresponding to the peak \ch{AlO} values within the resolved boundary layer. The current method avoids explicit resolution of the boundary layer by estimating the film properties from surface and ambient conditions,  and is therefore readily applicable to Lagrangian simulations of aluminum particles. {Without further calibration, the optimized correlation was also
evaluated for the {\SI{210}{\micro\metre}} spherical particle, predicting $\langle \dot{m}^{\prime\prime}\rangle{=}
\SI{0.705}{\kilogram\per\metre\squared\per\second}$,
compared with
\SI{0.729}{\kilogram\per\metre\squared\per\second}
from the boundary-layer resolved simulation, {which is} a deviation of {only} 3.24~\%. }


\subsubsection{Mass transfer modeling with oxide cap}
\label{subsubsec:mdot_cap}
Since the inclusion of an oxide cap alters the exposed surface area of the aluminum particle {and changes the flow and mixing field around the particle}, the overall mass evaporation rate, $\dot{m}$, calculated from $\langle \dot{m}^{\prime\prime} \rangle$ using the results of the boundary-layer resolved simulations described in Sec.~\ref{subsec:theory_resolved_particle}, is evaluated to enable a fair comparison among different oxide caps and ${\rm Re_p}$. {It is noted that the mass of aluminum remains constant for different $f_{\rm ox}$.} The results are shown in Fig.~\ref{fig:m_Red} using markers. The filled and unfilled markers represent the front and rear caps, respectively. It is noted that, for all caps, an increase in ${\rm Re_p}$ corresponds to a higher $\dot{m}$, despite a reduction in $\langle T_{\rm s} \rangle$ at higher ${\rm Re_p}$. The presence of an oxide cap inhibits the evaporation ($\dot{m}$ decreases from that of $f_{\rm ox} = 0$), and this inhibition becomes more pronounced as $f_{\rm ox}$ increases. In particular, a rear cap leads to a stronger inhibition of evaporation, especially at high $\rm Re_p$. The Lagrangian method proposed in Sec.~\ref{subsec:mdot_sphere} for spherical burning aluminum particles is applied with Eq.~\ref{eq:lagrangian_m_modified}, using the optimized Whitaker correlation and $\zeta=1$, following the common assumption that $\langle \dot{m}^{\prime\prime}\rangle$ remains unchanged when an oxide cap is involved. 
The results are observed to agree only with {boundary-layer resolved simulations for the }rear cap. The assumption that $\langle \dot{m}^{\prime\prime}\rangle$ remains unchanged is invalid for the front-cap configuration. Therefore, a different $\zeta$ is required. Here, a factor as a function of $\beta_{\rm ox}$ and $\rm Re_f$ is proposed:
\begin{equation}\label{eq:lagrangian_mdot_correct}
\zeta = 1+0.0042\,\mathrm{Re_f}+0.0102\,\beta_\mathrm{ox}.
\end{equation}
{This} expression for $\zeta$ performs well for the front-cap configuration, with the predictions shown by solid lines. For $\beta_\mathrm{ox}=0$ in the quiescent environment, the current expression provides a $\zeta$ close to the theoretical value 1. The corrected model was additionally evaluated for the front-cap {\SI{210}{\micro\metre}} particle with $f_{\rm ox}=0.3$, at
$\mathrm{Re_p}=6.38$, yielding an
$\dot{m}$ of $8.454\times10^{-8}$\,\si{\kilogram\per\second},
compared with
$8.012\times10^{-8}$\,\si{\kilogram\per\second}
from the boundary-layer resolved simulation, a deviation of only $5.52\%$. 
Wang et al.~\cite{wang2026alo} experimentally observed a stabilized asymmetric configuration with a downstream oxide cap following droplet coalescence, while Brykov et al.~\cite{brykov2023drag} reported that the preferred cap orientation depended on the aluminum mass fraction. Further investigation is therefore needed to establish the preferred orientation across the cap sizes and combustion conditions considered here, particularly given the pronounced orientation effect on evaporation at \(f_{\rm ox}=0.3\). 

\begin{figure}[h!]
\centering
\includegraphics[width=185pt,trim=0pt 0pt 0pt 0pt, clip]{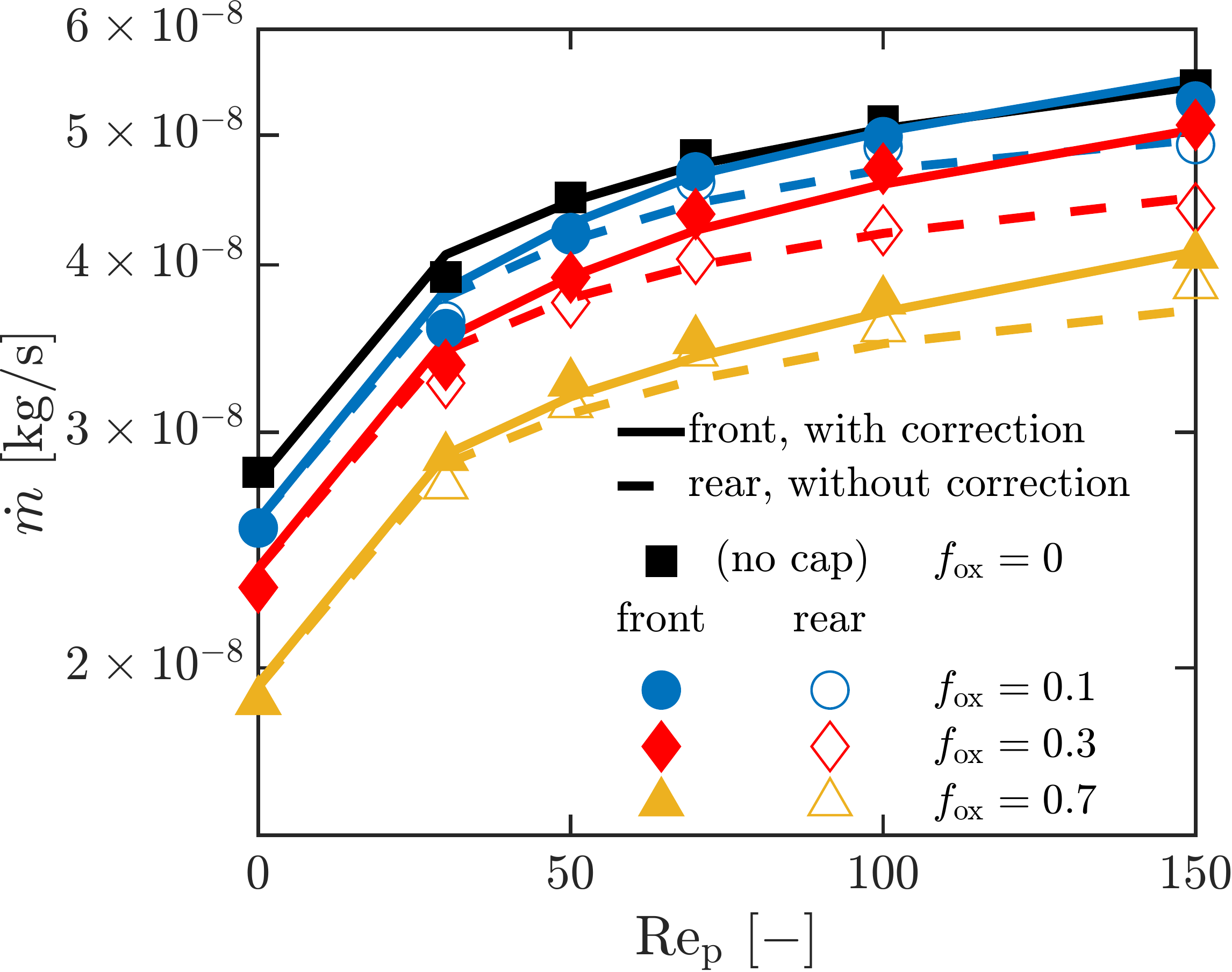}
\vspace{0 pt}
\caption{\footnotesize Comparison of {boundary-layer resolved} mass evaporation rate (markers) and predictions using Lagrangian correlations (lines) for different $\rm Re_p$ and oxide cap sizes, positioned both upstream and downstream of the aluminum particle.}
\vspace{-5 pt}
\label{fig:m_Red}
\end{figure}

{
Figure~\ref{fig:mdot_theta} shows the local evaporation characteristics of the front- and rear-cap configurations as a function of $\theta$. The local evaporation mass flux $\dot{m}^{\prime\prime}_{\theta}$ and the local evaporation rate per unit polar angle $\dot{m}_{\theta}$ are analyzed. The two quantities distinguish the intensity of local evaporation from its contribution to the total evaporation rate. For the front-cap configurations, $\dot{m}^{\prime\prime}_{\theta}$ increases substantially near the cap-aluminum intersection line. This enhancement occurs in the angular region where elevated $\chi$ was identified in Sec.~\ref{subsubsec:mixing}. A higher $\mathrm{Re_p}$ corresponds to a more pronounced enhancement. In contrast, the rear-cap configurations retain $\dot{m}^{\prime\prime}_{\theta}$ closer to that of the spherical particle over the majority of the upstream exposed aluminum surface, consistent with their similarity in the upstream mixing characteristics. A weaker local enhancement occurs near the cap-aluminum junction, where aluminum, alumina, and gas meet. A similar trend is also observed in $\dot{m}_{\theta}$. Although both configurations exhibit local changes near the cap–aluminum junction, the model without an additional correction already provides good overall predictions for the rear-cap cases. The stronger enhancement in the front-cap cases supports the use of the additional correction.}

\begin{figure}[h!]
\centering
\includegraphics[width=190pt,trim=0pt 0pt 0pt 0pt, clip]{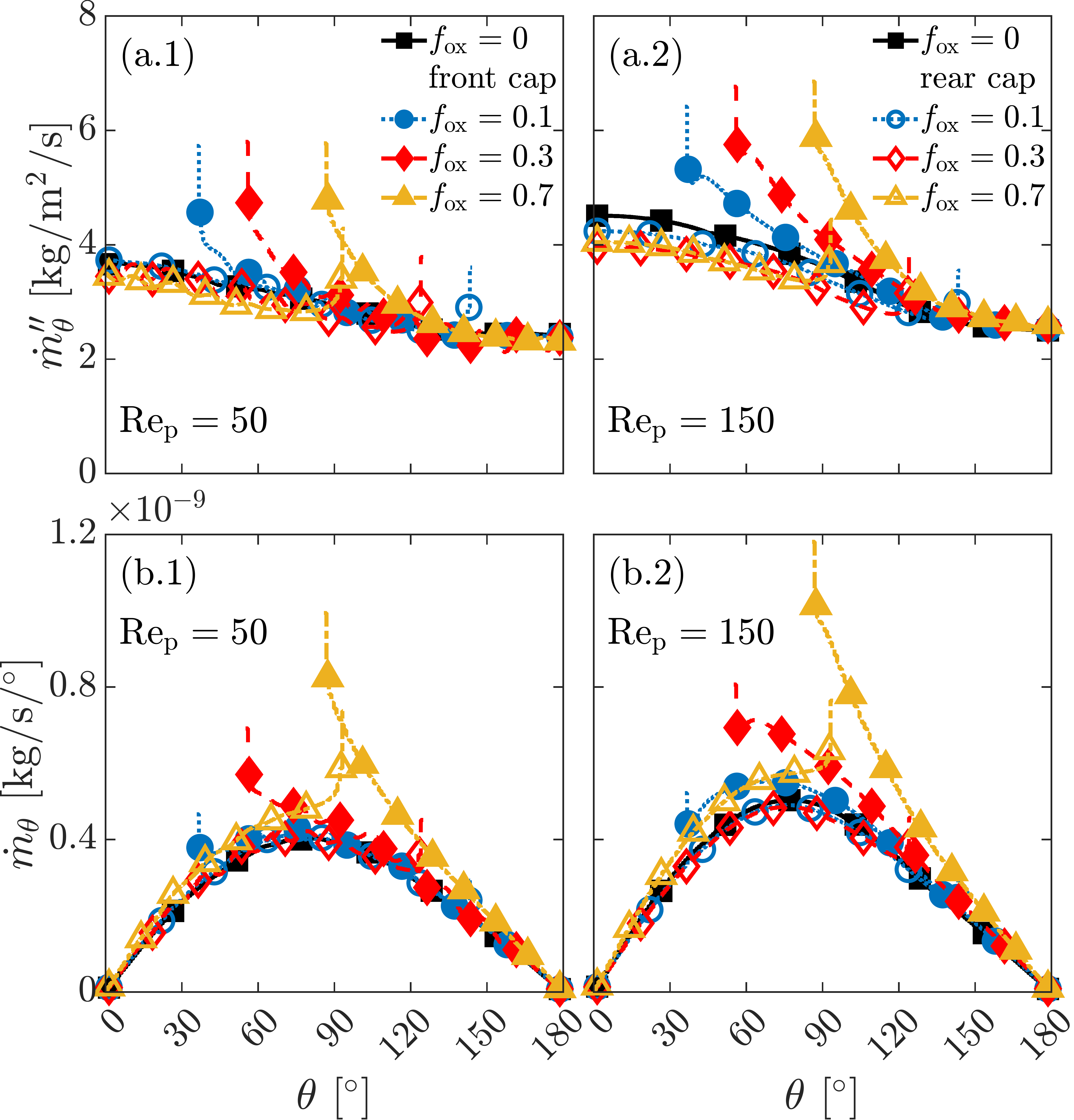}
\caption{\footnotesize Local surface properties as a function of $\theta$, which is illustrated in Fig.~\ref{fig:contour_cap}, at $\rm Re_p = 50 $ and $150$. (a) $\dot{m}^{\prime\prime}_{\theta}$ and (b) $\dot{m}_{\theta}$. }
\vspace{-10 pt}
\label{fig:mdot_theta}
\end{figure}

\subsection{Lagrangian modeling of drag force}
This section aims to accurately model drag force within the Lagrangian framework. The drag force and its pressure and viscous components were calculated from the results of boundary-layer resolved simulations following Eqs.~\ref{eq:drag} to \ref{eq:drag_tau}. From the perspective of Lagrangian modeling of the drag force, the drag coefficient, ${\rm C_D}$, is calculated using Eqs.~\ref{eq:drag} and \ref{eq:lagrangian_drag}, with the results from boundary-layer resolved simulations used for comparison. The effects of an oxide cap are studied.

\subsubsection{Drag force modeling without {oxide} cap}
\label{subsubsec:drag_spherical}
The drag force $\bm{F}_{\rm D}$ is first evaluated for spherical particles. It is found that both $\bm{F}_{p}$ and $\bm{F}_{\tau}$ increase with $\rm Re_p$, and the two components are comparable. {The Lagrangian modeling of $\bm{F}_{\rm D}$ depends on the $\rm C_D$ correlation, which is approximated by Eq.~\ref{eq:Cd}.} Figure~\ref{fig:Cd_Re} shows the inadequate prediction of the Schiller-Naumann correlation, compared with the results from boundary-layer resolved simulations. An improved version, $c_1 = 0.4$, $c_2 = 0.3780$, and $c_3 = 0.7344$, is therefore proposed. The new calibrated correlation noticeably improves the performance for $\rm Re_f<10$, where a more pronounced drag force is observed. It should be noted that the Reynolds number used in the calculation of $\rm C_D$ corresponds to the film state, which is approximated using the method proposed in Sec.~\ref{subsec:mdot_sphere}. Previous investigations have reported that the distinct boundary layer formed around a burning particle can alter the drag force, and have modified the $\rm C_D$ correlation by incorporating the Stefan flow velocity and the non-dimensional gaseous reaction rate~\cite{jayawickrama2019effect,zhang2020drag}. The currently proposed correlation remains consistent with the classical model through the employment of an approximation of film properties. The correlation between $\rm Re_f$ and $\rm Re_p$ is shown on the right-hand-side axis in Fig.~\ref{fig:Cd_Re}. The great difference between the two Reynolds numbers indicates that using $\rm Re_p$ to predict $\rm C_D$, instead of $\rm Re_f$, would lead to significant deviations and is therefore inappropriate. For fixed correlation coefficients, the film rule affects the predicted $\mathrm{C_D}$ only through $\mathrm{Re_f}$. For comparison, $\rm C_D$ is also plotted using $\rm Re_f$ calculated from the 1/3-rule, shown by unfilled markers. These markers deviate from both versions of the $\rm C_D$ correlation. Although a calibrated correlation is expected to capture the trend, no additional calibration is performed in the present study to maintain a consistent film rule.

\begin{figure}[h!]
\centering
\includegraphics[width=190pt,trim=0pt 0pt 0pt 0pt, clip]{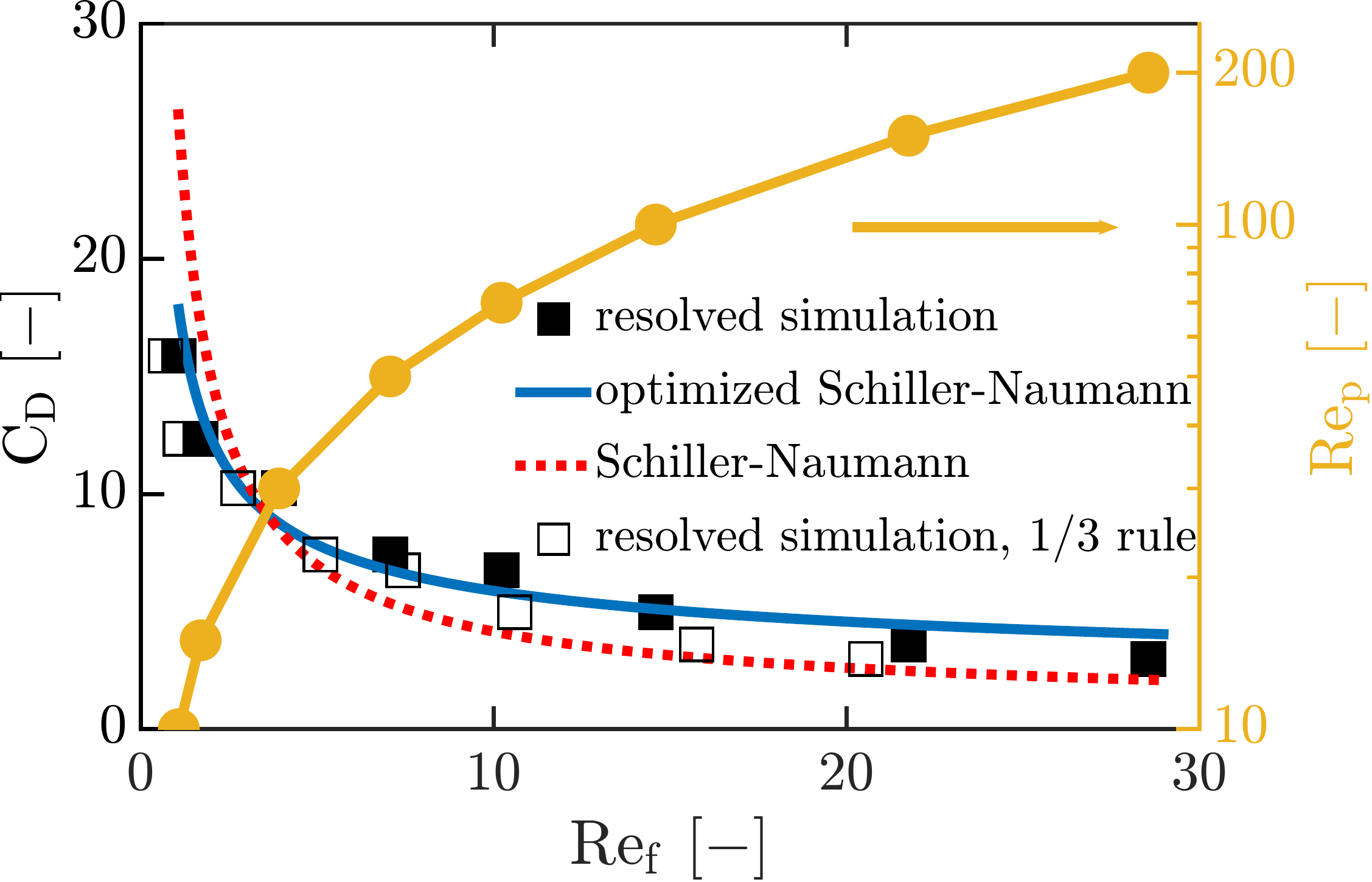}
\caption{\footnotesize Comparison of drag force coefficients obtained from Lagrangian correlations and boundary-layer resolved simulations. The correlation between $\rm Re_f$ and $\rm Re_p$ is shown on the right-hand-side axis. The correlation between the simulation results and the $\rm Re_f$ calculated from the 1/3 rule is plotted using unfilled markers.}
\vspace{-5 pt}
\label{fig:Cd_Re}
\end{figure}

\subsubsection{Drag force modeling with {oxide} cap}
\label{subsubsec:drag_cap}
{
In this section, the drag force coefficient, $\mathrm{C_D}$, of doublet aluminum particles with an oxide cap is evaluated from the boundary-layer
resolved simulations. Unlike $\dot{m}$, which is evaluated over the exposed aluminum surface, $\mathrm{C_D}$ accounts for the forces acting
on both the aluminum body and the oxide cap. Figure~\ref{fig:drag_non_spherical} compares the results with the optimized Schiller-Naumann correlation for spherical aluminum particles obtained in Sec.~\ref{subsubsec:drag_spherical}. Note that the comparison is presented as a function of $\mathrm{Re_f}$, which is used in the Lagrangian correlation of $\mathrm{C_D}$. At each $f_{\rm ox}$,
the front- and rear-cap cases correspond to the same prescribed $\mathrm{Re_p}$. Nevertheless, their resulting $\mathrm{Re_f}$ can 
differ due to the different flame characteristics. This difference is more pronounced for high $\mathrm{Re_p}$. 

For both cap positions, $\mathrm{C_D}$ decreases with increasing $\mathrm{Re_f}$, following the overall trend for spherical particles. Although the correlation underpredicts $\mathrm{C_D}$ for the rear-cap configuration at $f_{\mathrm{ox}}=0.1$ and low Reynolds numbers, it generally provides an approximate description of the capped-particle results. Note that the present $\mathrm{C_D}$ calculation uses
$A_{\rm c}=\pi r_1^2$ as the reference cross-sectional area, where $r_1$ is the radius of the spherical aluminum body. Alternative reference areas, including the cross-sectional area of a spherical particle with the same total volume as the doublet particle and the actual projected area of the doublet particle normal
to the inflow, were also examined. These alternative normalizations did not show comparable overall agreement with the optimized spherical correlation to the currently adopted $A_{\rm c}$, and the detailed results are omitted for brevity.
}

\begin{figure}[h!]
\centering
\vspace{-5 pt}
\includegraphics[width=180pt,trim=0pt 0pt 0pt 0pt, clip]{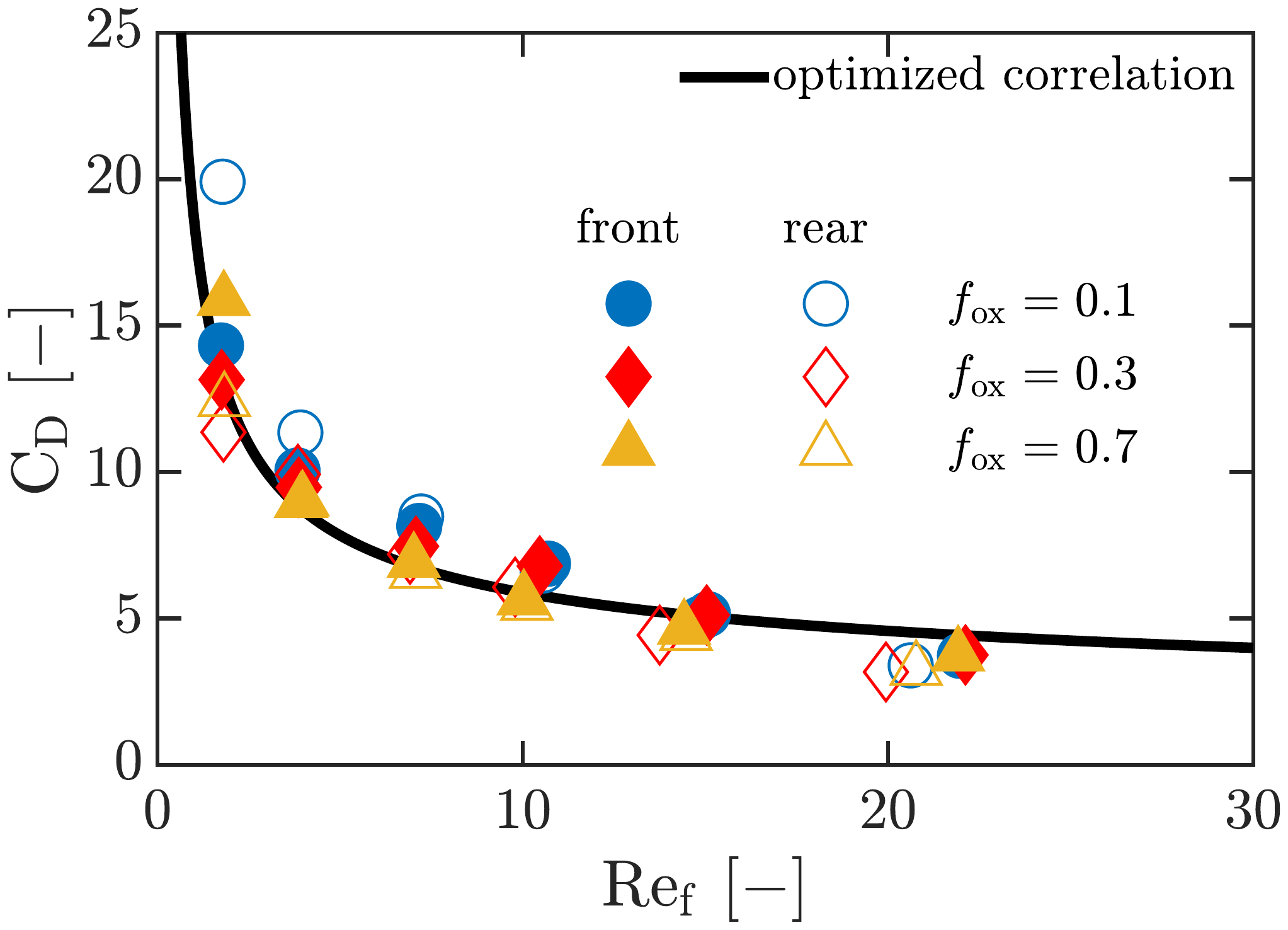}
\vspace{-5 pt}
\caption{\footnotesize Comparison of drag force coefficients obtained from the optimized Schiller-Naumann correlation (solid line) and boundary-layer resolved simulations (markers) for different $\rm Re_f$ and oxide cap sizes, positioned
both upstream and downstream of the aluminum particle.}
\vspace{-5 pt}
\label{fig:drag_non_spherical}
\end{figure}

{
Figure~\ref{fig:drag_components} presents the pressure contribution, $\mathrm{C}_{\mathrm{D},p}$, and the
viscous contribution, $\mathrm{C}_{\mathrm{D},\tau}$, to $\mathrm{C_D}$. The two contributions are comparable for all the simulated cases. At $\mathrm{Re_p}=15$ and $f_{\rm ox}=0.1$, the elevated
$\mathrm{C_D}$ of the rear-cap configuration is associated with increases in both contributions relative to the front-cap and spherical cases, with a larger increase in the viscous contribution.}

\begin{figure}[t]
    \centering
    \includegraphics[width=\columnwidth]{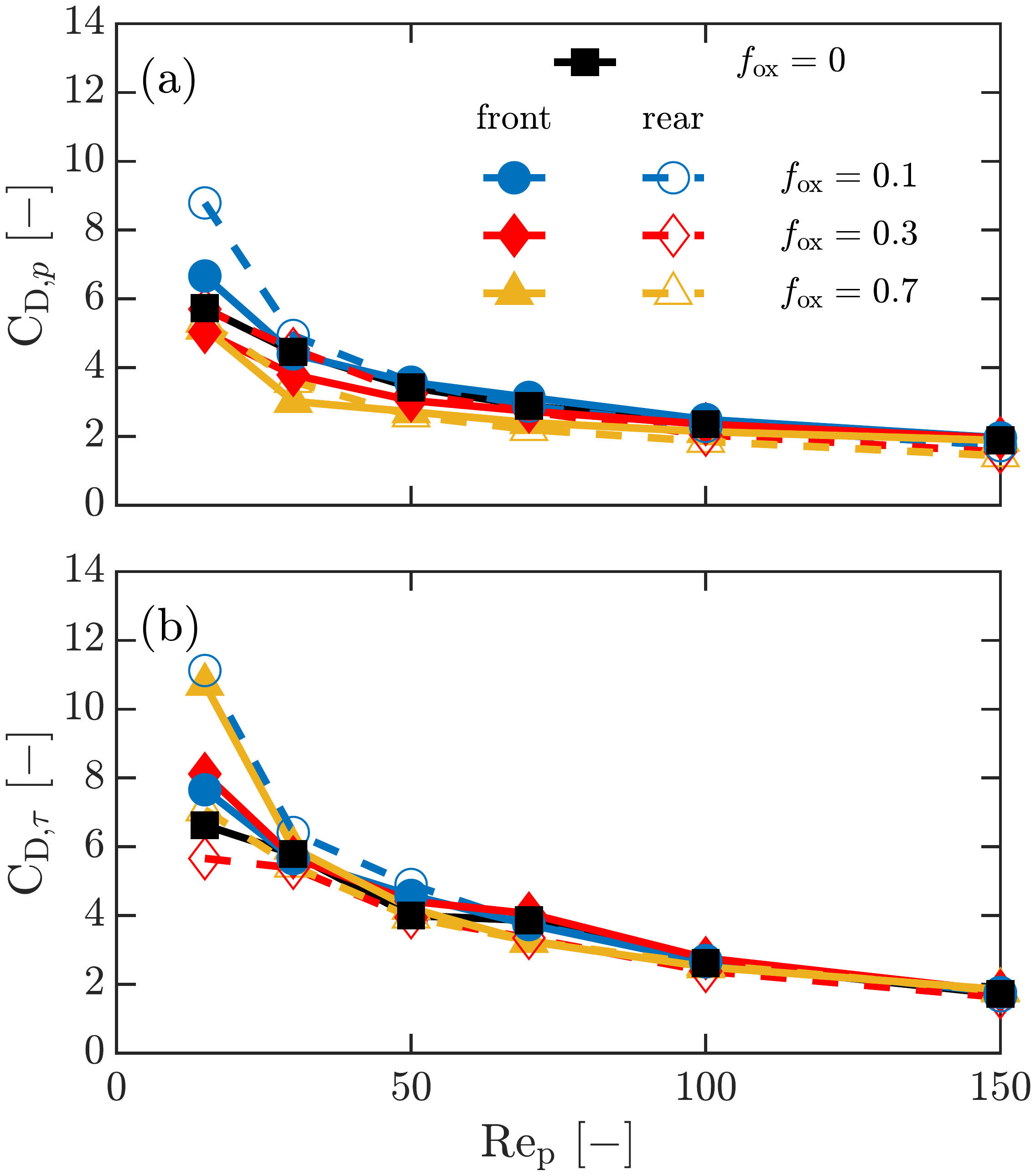}
    \vspace{-10 pt}
    \caption{
    Pressure and viscous contributions to $\mathrm{C_D}$
    as functions of $\mathrm{Re_p}$ for spherical particles and particles with front- and rear-positioned oxide caps:
    (a) $\mathrm{C}_{\mathrm{D},p}$ and (b) $\mathrm{C}_{\mathrm{D},\tau}$.
    }
    \vspace{-5 pt}
    \label{fig:drag_components}
    \end{figure}
\section{Summary}
Boundary-layer resolved simulations were performed to investigate the effects of oxide cap size and position on burning aluminum particles over a range of $\mathrm{Re_p}$.
{The computational framework was validated against the measurements of Bucher et al.~\cite{bucher2000detailed}, capturing the main features of the normalized
radial profiles of \ce{AlO} and condensed \ce{Al2O3}.}

{The oxide cap modifies both the interfacial transfer and the surrounding flame characteristics. Front caps enhance local evaporation near the junction between the cap and the aluminum body. In contrast, rear caps generally preserve the upstream mixing characteristics of spherical particles. Heat absorption by the rear cap contributes to weaker
chemical reactions in the surrounding gas. Variations in integrated heat release generally follow those in the evaporation rate. Alumina condensation contributes approximately half of the integrated heat release in the investigated cases.}


Based on the simulation results of spherical aluminum particles, the Lagrangian modeling framework was revised. The film state is approximated using a 0.55-rule for temperature, species mass fractions, and pressure, from which the remaining thermophysical properties are derived. {A simplified composition scheme enables film-state viscosity and fuel diffusivity to be pre-calculated as functions of temperature, and is expected to 
reduce the computational cost.} Using the film properties, optimized Whitaker and Schiller–Naumann correlations were developed for predicting the mass evaporation rate, $\dot{m}$, and the drag force coefficient, ${\rm C_D}$, respectively. Both show good agreement with the boundary-layer resolved simulations.


{The framework was further extended to doublet particles with oxide caps. Accounting for the exposed aluminum area provides good evaporation predictions for rear-cap configurations. For front-cap configurations, however, the assumption of unchanged surface-averaged evaporation mass flux is inadequate, particularly at high $\mathrm{Re_p}$. A correction depending on $\mathrm{Re_f}$ and the surface coverage $\beta_{\rm ox}$ was proposed, which improves the predictions. Using $A_{\rm c}=\pi r_1^2$ as the reference area, where $r_1$ is the radius of the spherical aluminum body and varies with the oxide cap size at {fixed mass of aluminum body}, the optimized spherical drag correlation generally provides an approximate description of the capped-particle results. The correlations were calibrated for the investigated quasi-steady conditions. Their applicability beyond this range requires further assessment.}

\section*{CrediT authorship contribution statement}
\textbf{Hesheng Bao}: Conceptualization, Software, Methodology, Writing – original draft. \textbf{Hongchao Chu}: Conceptualization, Methodology, Writing – review \& editing.  \textbf{Shibam Bose}: Writing – review \& editing. \textbf{Heinz Pitsch}: Supervision, Conceptualization, Writing – review \& editing. 

\section*{Declaration of competing interest}
The authors declare that they have no known competing financial
interests or personal relationships that could have appeared to
influence the work reported in this paper.

\section*{Acknowledgments}
The authors appreciate the funding by the German Federal Ministry of Education and Research under Research Unit AL-H2O-REAC. The authors gratefully acknowledge the computing time provided to them at the NHR Center NHR4CES at RWTH Aachen University (project number: p0024040). This is funded by the Federal Ministry of Education and Research, and the state governments participating on the basis of the resolutions of the GWK for national high performance computing at universities (www.nhr-verein.de/unsere-partner).

\section*{Declaration of the use of generative AI}

The authors used Codex (OpenAI) to refine author-written text. The authors reviewed and edited all affected content and take full responsibility for the published article.

\FloatBarrier

\bibliographystyle{cnf-num}
\bibliography{cnf-refs}

\end{document}